\documentclass[sigconf]{acmart}

\usepackage{epstopdf}
\usepackage{textcomp}

\usepackage{url}
\usepackage{graphicx}
\usepackage{amsmath,amssymb,amsfonts}
\usepackage{xcolor}
\usepackage{color}
\usepackage{booktabs,subcaption,dcolumn}

\usepackage{url}

\usepackage{algorithm}
\usepackage{algpseudocode}
\usepackage{arydshln}
\usepackage{mathtools}

\newcommand{\sysname}[1]{\textsc{#1}}

\newcommand{\bP}{{\bf P}}

\newcommand{\bU}{{\bf U}}
\newcommand{\bu}{{\bf u}}
\newcommand{\bV}{{\bf V}}
\newcommand{\bX}{{\bf X}}

\newcommand{\bY}{{\bf Y}}

\newcommand{\defeq}{\vcentcolon=}
\newcommand{\srank}{\text{srank}}
\newcommand{\rank}{\text{rank}}
\newcommand{\sw}{\textsc{Sw-dae }}
\newcommand{\fc}{\textsc{Fc-dae }}

\AtBeginDocument{%
  \providecommand\BibTeX{{%
    \normalfont B\kern-0.5em{\scshape i\kern-0.25em b}\kern-0.8em\TeX}}}

\copyrightyear{2020}
\acmYear{2020}
\acmConference[WWW '20]{Proceedings of The Web Conference 2020}{April 20--24, 2020}{Taipei, Taiwan}
\acmBooktitle{Proceedings of The Web Conference 2020 (WWW '20), April 20--24, 2020, Taipei, Taiwan}
\acmPrice{}
\acmDOI{10.1145/3366423.3380135}
\acmISBN{978-1-4503-7023-3/20/04}

\begin{document}

\title{Learning the Structure of Auto-Encoding Recommenders} 

\author{Farhan Khawar}
\orcid{0000-0001-9470-9827}
\affiliation{%
  \institution{The Hong Kong University of \\Science and Technology}
}
\email{fkhawar@connect.ust.hk}

\author{Leonard K.M. Poon}
\orcid{0000-0002-8394-1492}
\affiliation{%
  \institution{The Education University of \\Hong Kong}
}
\email{kmpoon@eduhk.hk}

\author{Nevin L. Zhang}
\orcid{0000-0002-4662-3217}
\affiliation{%
  \institution{The Hong Kong University of \\Science and Technology}
}
\email{lzhang@cse.ust.hk}


\begin{abstract}
Autoencoder recommenders have recently shown state-of-the-art performance in the recommendation task due to their ability to model non-linear item relationships effectively. However, existing autoencoder recommenders use fully-connected neural network layers and do not employ structure learning. This can lead to inefficient training, especially when the data is sparse as commonly found in collaborative filtering. The aforementioned results in lower generalization ability and reduced performance. In this paper, we introduce structure learning for autoencoder recommenders by taking advantage of the inherent item groups present in the collaborative filtering domain. Due to the nature of items in general, we know that certain items are more related to each other than to other items. Based on this, we propose a method that first learns groups of related items and then uses this information to determine the connectivity structure of an auto-encoding neural network. This results in a network that is sparsely connected. This sparse structure can be viewed as a prior that guides the network training. Empirically we demonstrate that the proposed structure learning enables the autoencoder to converge to a local optimum with a much smaller spectral norm and generalization error bound than the fully-connected network. The resultant sparse network considerably outperforms the state-of-the-art methods like \textsc{Mult-vae/Mult-dae} on multiple benchmarked datasets even when the same number of parameters and flops are used. It also has a better cold-start performance.  

\end{abstract}

\keywords{ Structure Learning, Collaborative Filtering, Sparse Autoencoder, Wide Autoencoder, Shallow Networks.}

\maketitle

\section{Introduction}

Collaborative filtering (CF) uses the past behavior of users to recommend items to users\cite{marlin2004collaborative,Koren2015,khawar2019modeling,khawar2019conformative}. This past behavior is given in the form of a user-item matrix, where each row represents a user and each element in the row indicates whether the user has consumed\footnote{Consumed may refer to bought, viewed, clicked, liked or rated, etc.} the corresponding item or not.

Recently, autoencoder recommenders \cite{liang2018variational,wu2016collaborative} have been introduced to learn the user/item representations in a non-linear fashion and have shown to outperform conventional methods \cite{liang2018variational,wu2016collaborative}. These existing methods use neural networks with fully-connected (FC) layers where each neuron of the previous layer is connected to all the neurons of the next layer. An FC network structure is general purpose and carries a high modeling capacity, therefore it is a good first choice for application in any domain. However, more often than not, each domain has a certain structure that can be exploited to design neural network architectures that are less general but nevertheless more suitable for that particular domain. For example, convolution neural networks exploit the spatial invariance property of images to reduce the network size and increase the performance compared to the FC network. Similarly, recurrent neural networks are another example that exploits the sequential nature of the text data. Surprisingly, existing autoencoder recommenders (or even other neural network recommenders) use FC networks and have not explored structure learning in general and the use of domain-specific information to learn the structure of the neural network in particular.

\begin{figure}
\includegraphics[width=1.0\columnwidth]{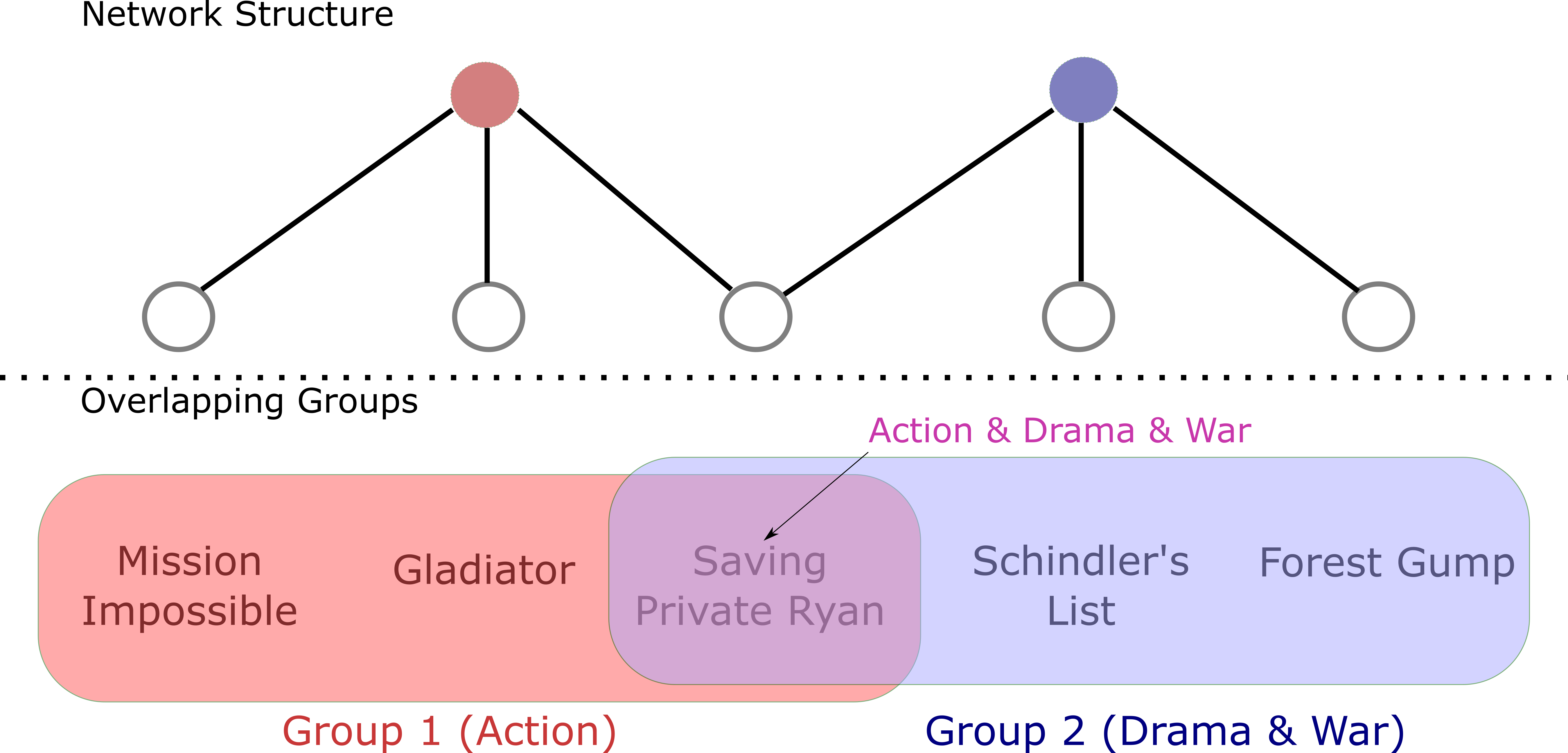}
\vspace{-3mm}
\caption{Sample clusters from the Movielens20M dataset are shown. Items are related to each other. Each item can be related to more than one group of items. Neural networks are built based on the overlapping groups of items. } \label{figure.groups}
\vspace{-.5em}
\end{figure}

The data in the CF domain has an inherent structure, that is, certain items tend to be more related to each other than to other items. As an example, consider the groups of items from the Movielens20M data shown in Figure~\ref{figure.groups}. We can see that groups of thematically similar items exist. Moreover, we also note that an item can belong to multiple groups, for example, the movie Saving Private Ryan is both an action movie and a drama \& war movie, so it is similar to movies in both groups. 

With this premise, the attempt of the FC layer to use each neuron to model the interactions of \emph{all} items can lead to modelling unnecessary/noisy item interactions which can harm the generalization ability of the autoencoder. This phenomenon of learning unnecessary parameters can be more detrimental when the data is sparse. Motivated by this, we propose to use structure learning techniques to determine the connections between the layers of the autoencoder recommender.

Common structure learning techniques like pruning connections with small weights~\cite{han2015learning} and using $\ell_1$-regularization on neuron activation~\cite{Jiang2013ANS} are a natural choice and can easily be adapted for recommendation. However, they first start with an overcomplete FC network and then contract the network by retaining useful connections. While this approach is promising as it will be better than asking each neuron to model all item interactions, it does not take advantage of the prior information available to us regarding the tendency of items to be related to each other. Also, by staring with an FC network these approaches rely on the neural network's ability to decide during training which connections are unnecessary. Instead, a better approach would be to use a disjoint two-stage method, where we first fix the network structure and remove the ``unnecessary'' connections and then train only these connections. This can be thought of as pointing the auto-encoder in the right direction before beginning the training. Thus, in a disjoint two-stage method, the first stage acts as prior that incorporates our domain knowledge about item groups and fixes the structure and the second stage trains only these connections of the network. This enables the network to train efficiently and have both a higher validation accuracy and a higher test accuracy compared to the alternatives.

In this paper, we propose a simple two-stage scheme to incorporate this domain knowledge into the autoencoder structure by presenting a structure learning method that determines the connectivity structure of the autoencoder based on item groups. Given the fact that some items tend to be related to each other and many items tend not to be related to each other, we can introduce a neuron for each group of related items. These neurons model the interactions between the related items and, unlike the FC networks, do not model the unnecessary interactions between unrelated items. We first find \emph{overlapping} groups of related items from the data and for each group, we introduce a neuron to model their interactions. The same structure is inverted and used for the decoder. Then, in the second step, we train this autoencoder to learn the weights of these connections.

Building a network structure in such a fashion allows the connectivity to be sparse as each item is only connected to a subset of neurons of the next layer. As a result, the number of parameters is reduced and the model has a smaller memory footprint than the corresponding FC networks. Since our structure is sparse we can make it wider compared to the FC network in the same memory footprint. Having a wide structure allows us to use more neurons to model the interactions of items. As a result, each neuron models the interaction of a few closely related items. We name the final structure as a sparse and wide (\textsc{Sw}) network.

We demonstrate the benefit of the \textsc{Sw} structure by using a \emph{denoising autoencoder} (\textsc{Dae}) \cite{vincent2008extracting}. FC and non-wide \textsc{Dae}s have been used successfully in the past for the recommendation task\cite{liang2018variational,wu2016collaborative}, and have shown state-of-the-art performance\cite{liang2018variational}. By utilizing the item group structure of the data to make the network sparse and wide, the \textsc{Sw-dae} is able to considerably outperform existing methods on several benchmarked datasets. We also show that for the recommendation task, \textsc{Sw-dae} outperforms other common methods of neural network structure learning. Also, it exhibits superior performance compared to the state-of-the-art baseline in the cold-start scenario.

The main contributions of this paper are:
\begin{itemize}
    \item We introduce the idea of structure learning for the recommendation task and show that by incorporating existing structure learning techniques we can outperform the state-of-the-art deep learning recommenders.
    \item We then present a simple two-stage technique of learning the autoencoder connectivity structure based on item groups. Which, as we show, is better for the recommendation task than existing structure learning techniques.
    \item We demonstrate that this performance gain is due to the lower spectral norm of the weight matrices and hence a lower generalization error bound.
    \item Via empirical evaluation, we show that \sw exhibits the state-of-the-art performance even when it uses the same number of parameters/flops as the best baseline. Moreover, it is also superior in cold-start performance.
\end{itemize}

\section{Related Work}
\paragraph{De-noising Autoencoders} 
Autoencoders can be seen as a special case of feed-forward neural networks that attempt to recreate the input at their output. This is done by learning the hidden encoding of the input and using this encoding to recreate the input at the output. De-noising autoencoders \cite{vincent2008extracting} are a type of autoencoders that receive the input data in a \emph{noisy} form and attempt to recreate the original \emph{clean} data at their output.

\paragraph{Autoencoders and Neural Networks for Recommendation} 
Autoencoders have been used in the past for recommendations. Two such methods that have received attention are \textsc{Mult-vae} \cite{liang2018variational} and \textsc{Cdae} \cite{wu2016collaborative}. In \cite{wu2016collaborative} the authors extend the standard denoising autoencoder by using a specific neuron to model each user, whereas in \cite{liang2018variational} the authors introduce variational autoencoders (VAE) with the multinomial likelihood (\textsc{Mult-vae}) and a partially regularized objective function for recommendation. The \textsc{Mult-vae} represents the state-of-the-art performance on large scale real-world datasets. 

Another popular methods that extends linear factor methods for recommenders is neural collaborative filtering (\textsc{Ncf}) \cite{he2017neural}. \textsc{Ncf} introduces non-linear interactions between user and item latent vectors via a neural network. The number of parameters in \textsc{Ncf} increases linearly with the number of users and items and this can lead to degraded performance on larger datasets common in CF.

Unlike \textsc{Sw-dae}, all the aforementioned neural network based methods use fully-connected networks that do not incorporate the item group structure in their networks.
\paragraph{Structure Learning and Sparsity} 
Contraction approaches have been proposed to learn the structure and introduce sparsity in the neural network structure. They start by a larger than required network and then either prune the connections/neurons or introduce a penalty that forces the network to be sparsely activated. In \cite{han2015learning} a popular method of pruning is presented that prunes all the network connections with weights less than a predefined threshold. For making the de-noising autoencoder sparsely activated,  in \cite{Jiang2013ANS} an $\ell_1$ penalty is introduced on the activation of the hidden layer.  
Unlike \textsc{Sw-dae}, both these approaches of introducing learning the connectivity structure start with a complex model as the input and do not explicitly model the cluster structure.
\paragraph{Clustering} 
Clustering has been traditionally used for recommendations~\cite{ungar1998clustering}. User and item-based CF methods have high computational and space complexity. Clustering algorithms like K-means and hierarchical clustering have been applied to CF to improve the computational efficiency of these methods~\cite{Aggarwal:2016:RST:2931100,Amatriain2015}. But such methods generally fail to capture global patterns in data and trade accuracy for scalability~\cite{Amatriain2015,sarwar2002recommender,o1999clustering}. Other CF methods have also relied on clustering to aid in recommendation by clustering users/items based on the side information commonly found in heterogeneous networks~\cite{yu2014personalized}. Yet other methods like~\cite{Wu:2016:CIC:2835776.2835836,heckel2017scalable,khoshneshin2010incremental} have used co-clustering to get user/item latent factors. However, none of the above methods have used the item cluster information as a structural prior for the neural network and all of these methods exhibit inferior performance compared to the state-of-the-art deep-learning based methods.

\section{Learning the Sparse Structure}

The proposed method for learning the connectivity structure between two layers has the following steps: (i) group the input variables (items) into overlapping clusters, (ii) for each overlapping group introduce a neuron, (iii) learn the parameters of these neurons. 
We now describe each part of the structure learning procedure in detail.

\subsection{Getting the Overlapping Item Groups}
	\begin{figure}[	]
	\centering
		
		\setlength{\fboxrule}{0pt}
		\fbox{
		\includegraphics[scale=0.34]{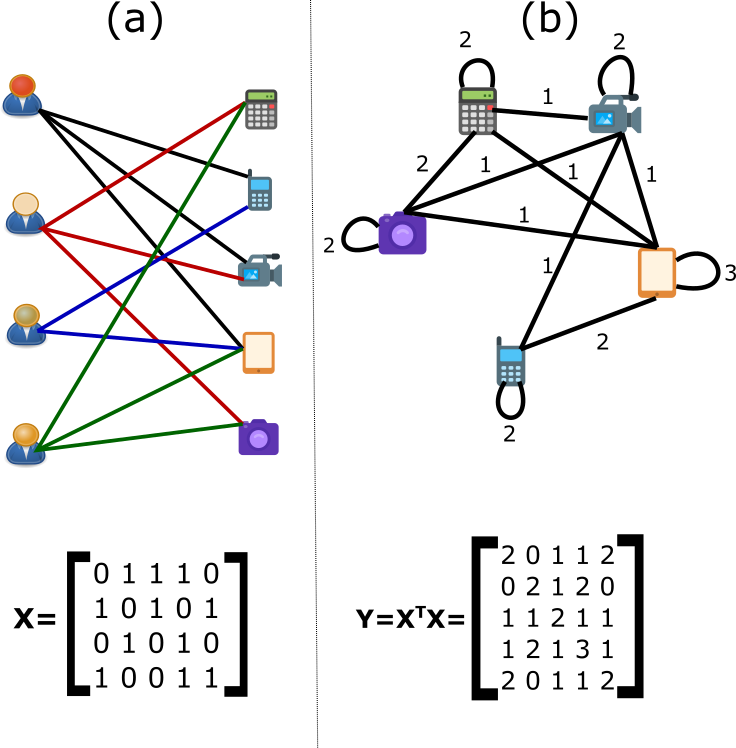}
		}
		\vspace{-7mm}
		\caption{A toy user-item bipartite graph and its corresponding user-item input matrix $\bX$ are shown in (a). It contains 4 users and 5 items. The corresponding item-item graph and its adjacency matrix $\bY$ are shown in (b).}
		\label{figure.graph}
		\vspace{-.5em} 
	\end{figure}

In this section, we consider 
the problem of partitioning the input variables into $K$ overlapping groups. This involves two steps: (1) getting item embeddings and (2) grouping the items based on these embeddings.

\subsubsection{Getting the Item Embeddings}
We would like to obtain item embeddings that (a) preserve the relationships between items and (b) discount for the effect of popular items. To achieve this we first use the user-item feedback matrix to obtain an item-item graph, then construct a graph laplacian that embodies our requirements, and finally get a low-dimensional representation of this laplacian to get our low-dimensional item embeddings.

Let $\bX$ be the $n \times m$ input matrix with $n$ users and $m$ items. We can view this matrix as a user-item bipartite graph, as shown in Figure \ref{figure.graph}(a), where each edge denotes that a user consumed an item. Since this is a bipartite graph, the items are related to each other via users. To get the item embeddings we first transform this into an item-item graph and then get the embedding of each node. This new item-item graph $G$ (shown in Figure \ref{figure.graph}(b)) can be easily obtained from the input matrix as its adjacency matrix is defined as $\bY = \bX^T\bX$.

Each edge in $G$ now denotes the existence of a co-occurrence relationship between items. In addition, $G$ is a weighted item-item graph where the weight of the edge denotes the strength of co-occurrence between items. This weight is equal to the number of times the two items were co-consumed by the users. From Figure \ref{figure.graph}(b) we can see that the item-item graph $G$ involves fewer nodes (items only) and the corresponding adjacency matrix representation is less sparse than the original bipartite matrix $\bX$. These observations are true in general and are the reason we chose to transform the input graph to $G$.

Given the item-item graph $G$ we would like to obtain node embeddings that (a) preserve the relationships between nodes in graph i.e., node embeddings should be influenced more by neighboring nodes with large edge weights and (b) account for the variation of node degrees in the graph i.e., a popular item will have a high co-occurrence with most items and may influence their neighbors more. Both these requirements can be fulfilled if we consider $\bX^T\bX$ as the adjacency matrix of $G$ and form the graph laplacian as:
\begin{equation}
\Delta = D^{-1/2} \bX^T\bX D^{-1/2}= D^{-1/2} \bY D^{-1/2},
\label{equation:laplacian}
\end{equation}
where, $D$ is the diagonal degree matrix and $D_{ii}= \deg(v_i) =\sum_j \bY_i$. $\bY_i$ is the $i-th$ row of $\bY$ and $\deg(v_i)$ is the degree of node $i$.

The elements of $\Delta$ are given by:
\begin{equation}
\Delta_{ij} := \begin{cases} \frac{\left|\mathbf{x}_i\right|}{\deg(v_i)} & \mbox{if } i = j \mbox{ and } \deg(v_i) \neq 0\\ \frac{\mathbf{x}_i.\mathbf{x}_j}{\sqrt{\deg(v_i)\deg(v_j)}} & \mbox{if } i \neq j \mbox{ and } v_i \mbox{ is adjacent to } v_j \\ 0 & \mbox{otherwise}. \end{cases}
\label{equation:laplacianexplain}
\end{equation}
Where, $\mathbf{x}_i$ denotes the $i-th$ column of $\bX$.  

To see how this laplacian fulfills our requirements, consider each row of the laplacian which corresponds to an item:
\begin{equation}
\Delta _i = \sum_{j:(i,j) \in \mathcal{E} } \frac{\mathbf{x}_i.\mathbf{x}_j}{\sqrt{\deg(v_i)\deg(v_j)}},
\end{equation}
where, $\mathcal{E}$ is the edge set of $G$. We see that the relationship between item $i$ and $j$ is proportional to their co-occurrence but inversely proportional to their popularity (node degrees). Thus, if we consider $\Delta_i$ to be an embedding of item $i$ then both our requirements (a) and (b) are fulfilled. 

However, $\Delta _i$ is an $m$-dimensional vector which can be large and as such we are interested in a low-dimensional embedding. To get this low-dimensional vector we perform the eigenvalue decomposition of $\Delta $ and pick the top $F$ eigenvectors. Briefly, the reason for picking the eigenvectors is that the eigenvectors of $\Delta$ can be seen as a smooth basis of the graph $G$ (i.e., corresponding to minimum Dirichlet energy) and smoothness in the frequency domain corresponds to spatial localization in the graph\cite{osting2014minimal,zosso2015dirichlet}. Thus, each eigenvector can be seen as partitioning $G$. We refer the reader to \cite{osting2014minimal,zosso2015dirichlet} for more details on the relationship between smoothness and spatial localization.

Having found the Laplacian $\Delta$, our task has now reduced to finding the eigenvectors of $\Delta$. First, the eigendecomposition of $\Delta= \bV \Lambda \bV^T$ is performed, where the columns of $\bV$ are the eigenvectors and $\Lambda$ is a diagonal matrix that contains the corresponding eigenvalues. Second, the largest $F$ ($F \ll m $) eigenvalues retained, since we want a low dimensional embedding, where $F$ is a user-provided parameter. Denote this matrix by $\bV_F$. The rows of $\bV_F$ represent the item embeddings in this low-dimensional space.

\subsubsection{Grouping}
After getting the item embeddings we turn our attention to grouping the items to obtain overlapping groups. We first start by clustering the items based on their embeddings using K-means\footnote{Any distance-based clustering can be used, we leave this exploration as future work.}. This gives us $K$ cluster centroids. Since each item can belong to more than one item clusters we connect each item to its $R$ ($R < K$) nearest centroids. This results in a simple and scalable approach to obtain the $K$ overlapping clusters.

The complete procedure is given in Algorithm \ref{alg.itemcluster}. First compute $\Delta$ using Equation \ref{equation:laplacianexplain} and compute the top $F$ left  eigenvectors $\mathbf{v}_1$, \ldots, $\mathbf{v}_F$ of $\Delta$ with the largest eigenvalues. Second, stack them column-wise to form the matrix $\bV_F$. Third, normalize each row of the matrix $\bV_F$ so that the sum is 1. Following \cite{ng2002spectral} this is done as a preprocessing step for clustering. Fourth, regard the rows of $\bV_F$ as points in the $F$-dimensional Euclidean space and use the K-means algorithm to get $K$ cluster centroids. Note that each row of $\bV_F$ corresponds to an item. In practice an item can belong to multiple clusters, therefore in step five, we connect each item to $R$ ($R < K$) nearest centroids. A partition of the items into $K$ overlapping clusters is therefore obtained.

\begin{algorithm}[]

\begin{description}

\item[Inputs:]  $\bX$ --- an $n \times m$ user-item matrix, $K$ --- number of clusters, $R$ --- degree of overlap, $F$ --- number of singular vectors.
\item[Outputs:]  $\;\;\;K$ overlapping item clusters.
\end{description}
\begin{algorithmic}[1]
\State Compute the $m \times m$ Laplacian matrix of $G$ using Equation \ref{equation:laplacianexplain}.
\State Find the $F$ largest eigenvectors of $\Delta$ via Lanczos Bidiagonalization: $\left \{\mathbf{v}_1 \dots \mathbf{v}_F\right \} =$ evd$(\Delta,F)$. \label{alg.svd}
\State Stack the eigenvectors column wise to form the matrix $\bV_F = [\mathbf{v}_1\dots \mathbf{v}_F]\in\mathbb{R}^{m \times F}$. \label{alg.stack}
\State Normalize $\bV_F$ row-wise such that each row sums to one. \label{alg.norm}
\State Run K-means on the rows of $\bV_F$ (i.e. the items) to get $K$ cluster centroids. \label{alg.kmenas}
\State Associate each item with $R$ nearest centroids to get $K$ overlapping item clusters.
\end{algorithmic}
\caption{\sysname{ItemGrouping}($\bX$, $K$, $R$, $F$)}\label{algo1}
\label{alg.itemcluster}
\end{algorithm}

\subsubsection{Discussions}
We compute the eigenvalue decomposition of $\Delta$ with the Lanczos bidiagonalization algorithm \cite{baglama2005augmented}. The Lanczos algorithm is fast for sparse matrices and its complexity is generally $O(n_{nz}F)$ \cite{cbai2000templates}, where $n_{nz}$ are the number of non-zero entries of $\Delta$. Since our laplacian is based on the co-occurrence matrix $\bY$, which is inherently sparse, getting the top $F$ eigenvectors is quite fast. This sparsity is the key to its scalability. Also, since each item is now an $F$-dimensional vector, the complexity of K-means would be $O(mlKF)$, where $l$ is the maximum number of iterations of K-means, and this is linear in the input size. We present more details on the scalability of Algorithm \ref{alg.itemcluster} in the section \ref{sec.itemgroupping}. 

Instead of forming the laplacian, we can get the eigenvectors of the co-occurrence matrix and project the items in the low dimensional space spanned by the eigenvectors. Since we don't form the laplacian, we can get the eigenvectors of $\bY$ directly from $\bX$. This is because $\bX=U\Sigma V^T$, and $\bY=\bX^T\bX=V\Sigma U^T U \Sigma V^T= V \Sigma^2 V^T$. Therefore, by performing the singular value decomposition of $\bX$ (using Lanczos bidiagonalization algorithm) we are able to get the eigenvectors of $\bY$. We can then follow a similar procedure to before, but now we operate on $\bX^T$ instead of the laplacian, perform its singular-value decomposition, and project the items in the low dimensional space and cluster them. The complete procedure is given in Algorithm \ref{alg.itemcluster2}. Unlike the laplacian, this method will have the disadvantage of allowing higher degree nodes to dominate more. However, since we operate on a more sparse matrix $\bX$, getting the top eigenvectors will be faster.

\begin{algorithm}[]

\begin{description}

\item[Inputs:] $\;\bX^T$ --- an $m \times n$ item by user implicit feedback matrix, $K$ --- number of clusters, $R$ --- degree of overlap, $F$ --- number of singular vectors.
\item[Outputs:]  $\;\;\;K$ overlapping item clusters.
\end{description}
\small
\begin{algorithmic}[1]
\State Find the $F$ largest left singular vectors and associated singular values of $\bX^T$ via Lanczos Bidiagonalization: $\left \{\bu_1\dots\bu_F, \sigma_1 \dots \sigma_F\right \} =$ svd$(\bX^T,F)$.
\State Stack the singular vectors column wise to form the matrix $\bU_F = [\bu_1\dots \bu_F]\in\mathbb{R}^{m \times F}$ and form the diagonal matrix $\Sigma_F$. 
\State Project the items in this space by $\bP=\bU_F \Sigma_F$\label{alg.project}.
\State Normalize $\bP$ row-wise such that each row sums to one. 
\State Run K-means on the rows of $\bP$ (i.e. the items) to get $K$ cluster centroids. 
\State Associate each item with $R$ nearest centroids to get $K$ overlapping item clusters.
\end{algorithmic}
\caption{\sysname{ItemGrouping2}($\bX^T$, $K$, $R$, $F$)}
\label{alg.itemcluster2}
\end{algorithm}

We note that Algorithm \ref{alg.itemcluster} is similar to spectral clustering\footnote{However, it is much faster due to operating on the sparse  co-occurrence matrix.} where it tries to embed items in a smooth space and Algorithm \ref{alg.itemcluster2} is like principal components analysis where the items are projected (line \ref{alg.project}) in the space spanned by the top eigenvectors of the co-occurrence matrix\footnote{PCA uses the dense co-variance matrix with expensive eigendecomposition.}. We compare the performance of these methods in section \ref{sec.itemgroupping}.

Another alternate procedure to get the overlapping clusters would be to cluster the columns of $\bX$. However, $\bX$ does not possess the desirable properties of $\Delta$. Also, due to the inherent sparsity of $\bX$, applying K-means directly on $\bX$ might lead to unsatisfactory performance. Finally, if applied directly on $\bX$, K-means will have a complexity of $O(mnKl)$. When $m$ and $n$ are both large this becomes computationally infeasible.    
\subsection{Building the Connectivity Structure} \label{build} 

\begin{figure}
\includegraphics[width=.88\columnwidth]{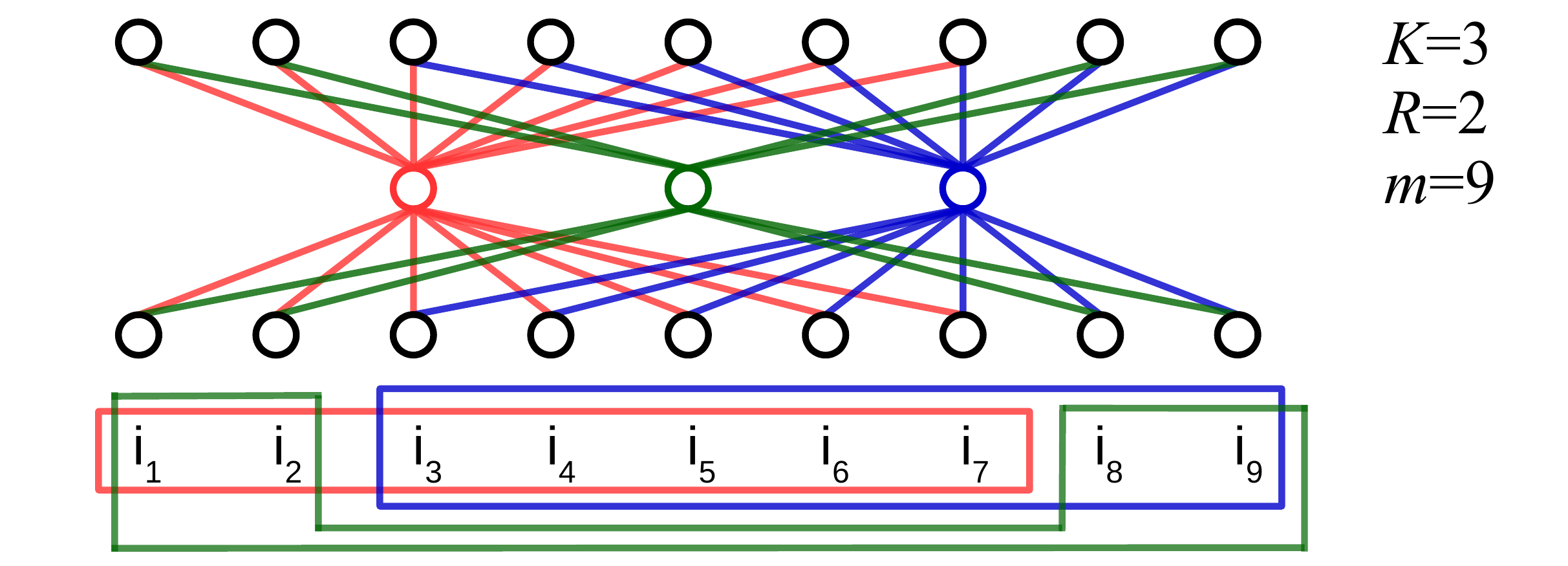}
\vspace{-3mm}
\caption{Illustration of network construction.  Suppose the items can be grouped into three overlapping clusters \{$i_{1}$, $i_{2}$, \ldots, $i_{7}$\}, \{$i_{3}$, $i_{4}$, \ldots, $i_{9}$\} and \{$i_{1}$, $i_{2}$, $i_{8}$, $i_{9}$\}. In the resulting structure, there is one neuron for each cluster (colored nodes).  Each neuron is connected to the input and output variables for all the items belonging to its cluster.} \label{figure.structure}
\vspace{-.8em}
\end{figure}

To determine the connectivity structure between two neural network layers we use the overlapping clusters of the items. Items in one cluster are more related to each other compared to items that are not in the same cluster. This is because each cluster represents items that are close to each other in the low-dimensional subspace. Therefore, we introduce a latent node (neuron) to model the interactions between them. This neuron connects to all the items in the overlapping cluster. Figure \ref{figure.structure} shows an illustration of the network structure.

By forming the structure in such a manner we ensure that (i) the connectivity is sparse as $R < K$, (ii) the interactions between related items are captured, and (iii) the aspect of an item being related to multiple item groups is also modeled. The intuition of forming such a connectivity structure is that an item is related to a few concepts (represented by neurons of the hidden layer) rather than to all concepts. 

\subsection{Learning Parameters}\label{learn}
Once the connectivity structure is learned we can use it to replace the FC layers in the denoising autoencoder. The denoising autoencoder consists of an input layer, a hidden layer, and an output layer. We use our learned structure to replace the FC connections between the input and the hidden layer, and the same connectivity structure is inverted to replace the connections between the hidden and the output layer. This results in a sparse and wide denoising autoencoder (\textsc{Sw-dae}).  

Let $\mathbf{x}=[x_i],i=1 \dots m$ denote the vector of input units, $\mathbf{h}=[h_j], j=1 \dots K$ denote the vector of the hidden units, $\mathbf{b}=[b_j]$ denote the bias vector of the hidden units, $\mathbf{W}=[w_{ij}]$ denote the weights of the connectivity structure such that $w_{ij}=0$ if input variable $x_i$ is not connected to the hidden variable $h_j$ and $\mathbf{W}^T$ denoting the transposition of $\mathbf{W}$. The \textsc{Sw-dae} then defines the probability distribution $p(\mathbf{h}|\mathbf{x})$ of the hidden units given the observed variables and the probability distribution $p(\mathbf{x}|\mathbf{h})$ of the observed variables given the hidden units as follows: 
\begin{equation}
\begin{split}
& p_{encoder}(\mathbf{h}| \mathbf{x}) = f(x)= \sigma(\mathbf{W}\mathbf{x}+\mathbf{b}),\\
& p_{decoder}(\mathbf{x} | \mathbf{h}) = g(h) =\sigma(\mathbf{W}'\mathbf{h}+\mathbf{b}'),
\end{split}
\end{equation}
where, $\sigma$ represents the sigmoid function and $\mathbf{W}'$ represents the 
decoder weight matrix which has the same connectivity structure as $\mathbf{W}^T$.

To learn the \textsc{Sw-dae} parameters i.e., the weights of the connections and biases of the neurons, we use the stochastic gradient descent with the denoising criterion \cite{vincent2008extracting}. Let $C(\mathbf{\tilde{x}}|\mathbf{x})$ be a random corruption process, which is a conditional distribution of the noisy sample $\mathbf{\tilde{x}}$ given the original data sample $\mathbf{x}$. Given the noisy sample $\mathbf{\tilde{x}}$ from $C(\mathbf{\tilde{x}}|\mathbf{x})$ as the input, the job of the \textsc{Sw-dae} is to learn the function to reconstruct the original data sample $\mathbf{x}$ i.e., learn the reconstruction distribution $p(\mathbf{x}| \mathbf{\tilde{x}})=p_{decoder}(\mathbf{x}|\mathbf{h}=f(\mathbf{\tilde{x}}))$. This can be achieved by performing stochastic gradient descent on the negative log-likelihood $-\log p_{decoder}(\mathbf{x}|\mathbf{h}=f(\mathbf{\tilde{x}}))$. This is equivalent to performing stochastic gradient descent to minimize the following expected loss \cite{Goodfellow-et-al-2016}:
\begin{equation}
{J} = -\mathbb{E}_{x \sim \hat{p}_{data}(\mathbf{x}) } \mathbb{E}_{\mathbf{\tilde{x}} \sim C(\mathbf{\tilde{x}}|\mathbf{x})} \log  p_{decoder}(\mathbf{x}|\mathbf{h}=f(\mathbf{\tilde{x}})),
\label{eqn.objective}
\end{equation}
where $\hat{p}_{data}$ is the empirical probability distribution of the data.
\subsection{Going Deeper}
The proposed method learns a single hidden layer autoencoder. In our experiments, this architecture gave the best results. 
However, for completeness, we also present how our structure learning technique can be used to go deeper. We propose a greedy layer-wise fashion to build a stacked \textsc{Sw-dae} structure and to get its pretrained weights. This is achieved by repeating the procedures of clustering the observed variables, building the connectivity structure and learning parameters until the desired depth is reached. 

The first step is to turn the hidden variables (neurons) to observed variables. This can be done by doing a forward pass for each data point $\mathbf{x}$, getting its hidden encoding and treating this as the new input vector. This results in the new data of dimension $n \times K$, where now we have $K$ observed variables. Then, in the second step, we perform the input variable clustering of Algorithm 1, followed by building the structure (section \ref{build}) and learning the parameters (section \ref{learn} ). This results in a new \textsc{Sw-dae}, which in the third step is stacked with the previous \textsc{Sw-dae}. We can repeat these three steps until the desired depth is reached.  

\begin{table}
\centering
\small
\caption{ Statistics of the datasets.} \vspace{-.5em}
\begin{tabular}{lcccc}
  \hline
   & \textbf{ML20M} & \textbf{Netflix} & \textbf{MSD}  \\
  \hline
  \# of users & 136,677 & 463,435 & 571,355  \\
  \# of items & 20,108 & 17,769 & 41,140  \\
  \# of interactions & 10.0M & 56.9M & 33.6M\\
  \% of interactions & 0.36\% & 0.69\% & 0.14\% \\
  \hline
  \# of val./test users & 10,000 & 40,000 & 50,000 \\
  \hline
\end{tabular}
\label{tab.dataset}
\vspace{-.5em} 
\end{table}

The \textsc{Sw-dae}s can be stacked by connecting the output of the encoder/decoder of each \textsc{Sw-dae} to the input of the successive \textsc{Sw-dae}'s encoder/decoder and copying the respective parameters. These copied parameters serve as pretrained weights. Formally, consider a stacked \textsc{Sw-dae} with $d$ hidden layers in its encoder/decoder, then the encoding $f(\mathbf{\tilde{x}})$ of the stacked \textsc{Sw-dae} is obtained by the successive encoding of each encoding layer $l=1 \dots d$ as follows:

\begin{equation}
\begin{split}
&\mathbf{z}^{l} = \mathbf{W}^{l-1}\mathbf{a}^{l-1} + \mathbf{b}^{l-1}, \\
&\mathbf{a}^l = \sigma(\mathbf{z}^l),
\end{split} 
\end{equation}
where 
%
$\mathbf{W}^l,\mathbf{a}^l$ and $\mathbf{b}^l $ denote the weight, activation and bias of the $l$-th layer respectively and $\mathbf{a}^0=\mathbf{\tilde{x}}$. The decoding $g(\mathbf{h})$ is also obtained by the successive decoding of the decoding layers but in the reverse order as follows:
\begin{equation}
\begin{split}
&\mathbf{z}^{d+l+1} = {\mathbf{W}^{d-l}}' \mathbf{a}^{d+l} + \mathbf{b}^{d+l},\\
&\mathbf{a}^{d+l}=\sigma(\mathbf{z}^{d+l}).
\end{split} 
\end{equation}

We note that $\mathbf{a}^d=f(\mathbf{\tilde{x}})=\mathbf{h}$ and $\mathbf{a}^{2d}=g(\mathbf{h})= p_{decoder}(\mathbf{x} | \mathbf{h}=f(\mathbf{\tilde{x}})$. Then the stacked \textsc{Sw-dae} can be trained using the objective of Equation \ref{eqn.objective} to fine tune the parameters.
	
\section{Empirical Study}
We perform an empirical evaluation of \textsc{Sw-dae} for the recommendation task. To perform the recommendation for a user, we input the user's binary consumption history (corresponding row of matrix $\mathbf{X})$ to the trained \textsc{Sw-dae}. Then we perform forward pass through the \textsc{Sw-dae} to get the output at the decoder before the softmax. The values in the output vector are treated as the score. We then rank the unconsumed items based on this score. 

\subsection{Datasets}
For a direct performance comparison, we use the same datasets, data splits (using the same random seed) and pre-processing steps as \cite{liang2018variational} during our evaluation. Details of the three datasets are below:
\begin{itemize}
\item Movielens20M (ML20M): is a dataset of users rating movies. The ratings were binarized and a rating value greater or equal to four was interpreted as a positive signal. Users who rated less than five movies were filtered out.
\item Netflix: is also a movie rating dataset from the Netflix prize{\footnote{\url{http://www.netflixprize.com/}}}. The ratings were binarized and the users with less than five movie ratings were removed and a rating values greater or equal to four was taken as positive.
\item Million Song Dataset (MSD) \cite{mcfee2012million}: is a dataset that contains the playing counts of users for songs. The counts were binarized and a user listening to a song was taken as a positive signal. Users who listened to less than twenty songs or songs that were listened to by less than two hundred users were removed. 
\end{itemize}

We use the strong generalization experimental setup \cite{liang2018variational,marlin2004collaborative} in our experiments. The datasets were split into training, validation and test users resulting in three subsets. The details of the splits are shown in Table \ref{tab.dataset}. The models were trained on the training users. A randomly chosen $80\%$ subset of the click history of the validation/test users was used to learn their necessary representations and the remaining $20\%$ of the click history was used for evaluation. 

\subsection{Metrics}
To get the ranked list of the unconsumed items of the validation/test users from \textsc{Sw-dae}, we ranked the items based on the un-normalized probability score at the decoder. We then used two top $R$ ranking based metrics, namely, Recall@$R$ and the normalized discounted cumulative gain NDCG@$R$ for evaluation. Formally, Recall@$R$ for a user $u$ is defined as:
\[
\textrm{Recall@}R(u, \rho) :=  \frac{\sum_{r=1}^R \mathbb{I}[\rho(r) \in I_u]}{\min(R, |I_u|)},
\]
where 
$\rho$ is the ranked list, $\rho(r)$ is the item at position $r$ in the list, $\mathbb{I}$ is the indicator function, and $I_u$ is the set of items consumed by user $u$. The term is the denominator ensures that Recall@$R$ has a maximum value of 1 which corresponds to raking all relevant items of the user in the top $R$ list.

The NDCG@$R$ is the DCG@$R$ divided by the best possible DCG@$R$. The best DCG@$R$ corresponds to ranking all the relevant items of the user at the head of the top $R$ list. Formally, DCG@$R$ is defined as:
\[
\textrm{DCG@}R(u, \rho) := \sum_{r=1}^R \frac{{\mathbb{I}[\rho(r) \in I_u]}}{\log(r+1)}.
\]
We note that unlike Recall@$R$ which gives equal importance to relevant items in the ranked list, NDCG@$R$ gives more importance to correctly ranked items at the head of the list than those lower down the list. 

\subsection{Experimental Setup} 
The hyperparameters and architecture were chosen based on the NDCG@$100$ performance on the validation set. The architecture of \textsc{Sw-dae} was symmetric for both the encoder and decoder. We used the sigmoid activation function as the non-linear activation function. For the corruption process, we applied a random dropout with a probability of $0.6$ at the input. During training, we also applied a dropout with a probability of $0.2$ at the hidden layer. We used a batch size of $500$ and trained using the Adam optimizer \cite{kingma2014adam} with a learning rate of $0.001$ for $100$ epochs. $K$ was searched in multiples of $1000$ up to the maximum value determined by the GPU memory for each dataset. For all our experiments, we set $F$ at $50$ and set $R$ to keep the network at $10\%$ sparsity (i.e., $R = 0.1K$) compared to the fully connected network.

\begin{table}[]
\fontsize{7}{7.0}\selectfont
\caption{The number of flops and parameters in millions. Even with the same number of parameters/flops \textsc{Sw-dae-p} provides considerable improvement in performance.}
\vspace{-3mm}
\label{tab:param}
\begin{tabular}{lclcc}
\hline
\textbf{ML20M}     & \textbf{Parameters (M)} & \textbf{Flops (M)} & \textbf{NDCG@100} & \textbf{\% Improvement} \\
\hline
\textsc{Sw-dae}   & 60.324                  & 120.645            & 0.442             & 3.76                    \\
\textsc{Sw-dae-p} & 24.129                  & 48.258             & 0.437             & 2.58\\
\textsc{Mult-vae} & 24.193                  & 48.738            & 0.426             & -\\
\hline
\multicolumn{4}{l}{\textbf{Netflix}}                                                  \\
\hline
\textsc{Sw-dae}   & 46.199                  & 92.3962            & 0.404             &4.66 \\
\textsc{Sw-dae-p} & 21.322                  & 42.644             & 0.398              & 3.11  \\
\textsc{Mult-vae} & 21.386                  & 43.124             & 0.386             & -\\
\hline
\multicolumn{4}{l}{\textbf{MSD}}                                                      \\
\hline
\textsc{Sw-dae}   & 74.052                  & 148.102            & 0.372             & 17.72 \\
\textsc{Sw-dae-p} & 49.368                  & 98.734             & 0.367             & 16.77\\
\textsc{Mult-vae} & 49.432                  & 99.214             & 0.316             & -\\
\hline
\end{tabular}
\vspace{-.5em}
\end{table}

\subsection{Baselines}
The following non-linear and linear state-of-the-art collaborative filtering methods were used as baselines in our experiments: 

\textsc{Mult-vae} \cite{liang2018variational}: is a non-linear recommender that uses a VAE with a partially regularized objective function and uses a multinomial loss function. The hyperparameters including the regularization parameter $\beta$ were set using the strategy of the original paper \cite{liang2018variational} that gave its best NDCG@$100$ performance. The Adam optimizer was used with a batch size of $500$. \textsc{Mult-dae} also uses the multinomial likelihood but uses a DAE instead of a VAE.

{{\textsc{Wmf}}} \cite{Hu:2008:CFI:1510528.1511352}: is a linear low-rank matrix factorization model. The weight of each consumption event was searched over the set $\{2, 5, 10, \\30, 50, 100\}$ and the size of latent dimension $K$ was searched over $\{100, 200\}$ based on the NDCG@$100$ performance on validation.

{\textsc{Slim}} \cite{ning2011slim}: is also a linear item based recommender which solves an $\ell_1$-regularized constrained optimization problem to learn a sparse item-item similarity matrix. The regularization parameters were searched over $\{0.1, 0.5, 1, 5\}$ based on the NDCG@$100$ performance on validation users.

{{\textsc{Cdae}}} \cite{wu2016collaborative}: enhances an autoencoder by adding a latent factor for each user. Unlike the DAE, this results in the number of parameters growing linearly with the number of users and items and makes it prone to overfitting. The settings of \cite{liang2018variational} were used to train the model. The weighted square loss and Adam with a batch size of $500$ was used for optimization.

We do not report results on \textsc{Ncf} since its performance was not competitive on the larger datasets used in this paper. This is consistent with the findings of \cite{liang2018variational}.

\section{Results}
In this section, we provide the quantitative results for our proposed approach compared with the baselines. We will look at the performance in order to answer the following:
\begin{itemize}
\item How do autoencoders with learned structure compare with baselines in terms of recommendation accuracy?
\item Is the \textsc{Sw-dae} gain due to more parameters or more neurons?
\item Does the item group structure in the autoencoder help?
\item How does the proposed structure learning approach compare with other structure learning techniques?
\end{itemize}

\subsection{Learned Structures Compared to Baselines}
In Table \ref{tab.results} we show the performance in terms of NDCG@$100$, Recall@$20$ and Recall@$50$. For each dataset, the structure learning methods are above the solid line and the baselines are below it. \textsc{Fc-prune} and \textsc{Fc-Reg} represent the use of pruning~\cite{han2015learning} and regularization~\cite{Jiang2013ANS} to learn the connectivity structure of the autoencoder respectively. We see that the standard structure learning methods outperform the baselines except for Recall@50 on ML20M. However, we see that the \textsc{Sw-dae} outperforms both the linear and non-linear baselines on all the datasets by a considerable margin. The largest performance improvement is observed on the sparsest MSD dataset. We conjecture that the addition of the cluster structure assists the network training when there are lower number of observations. 

\subsection{\textsc{Sw-dae} with the Same Number of Parameters} 
The results of the baseline neural network methods reported in Table \ref{tab.results} are based on the architecture with one or two hidden layers of $600$ or $200$ neurons since they gave the best validation NDCG@$100$ performance \cite{liang2018variational}. These architectures are FC and not as wide as \textsc{Sw-dae}, therefore, we also report the results of our method when the same number of parameters as the best baseline i.e., \textsc{Mult-vae} are used. We label this variant as \textsc{Sw-dae-p} and we can see that even with the same number of parameters \textsc{Sw-dae-p} outperforms the baselines. In Table \ref{tab:param} we show the comparison in terms of flops and parameters along with the percentage improvement in NDCG@100 of \textsc{Sw-dae/Sw-dae-p} over \textsc{Mult-vae}. We see that \sw is wider and has more parameters\footnote{These parameters are just 10\% of the parameters of the original FC network.} and flops but gives the best performance. However, \textsc{Sw-dae-p} with the same number of parameters and flops as \textsc{Mult-vae} is still able to considerably outperform it on all datasets.

\begin{table}
\small
\caption{Comparison between \textsc{Sw-dae} with various baselines on the test set. \textsc{Sw-dae} outperforms the baselines considerably on all datasets. \textsc{Slim} did not finish within a reasonable amount of time on MSD.}
\vspace{-3mm}
\begin{subtable}[]{\columnwidth}
\vspace{-2mm}
\caption{ML-20M}
\vspace{-2mm}
	\centering
\begin{tabular}{ l c c c c }
\hline
    & Recall@20 & Recall@50 & NDCG@100  \\
  \hline
    \textsc{Sw-dae}&\bf 0.410&\bf 0.549&\bf 0.442\\
    \textsc{Sw-dae-p}&0.406& 0.542& 0.437\\
  \hdashline
  \textsc{Fc-Prune-50}&0.399&0.534&0.431\\
  \textsc{Fc-reg}&0.399&0.535&0.431\\
  \hline
  \textsc{Mult-vae} &  0.395 &  0.537 &  0.426 \\
  \textsc{Mult-dae} & 0.387 & 0.524 & 0.419 \\
  
  \textsc{Wmf} & 0.360 & 0.498 & 0.386 \\
  \textsc{Slim} & 0.370 & 0.495 & 0.401\\
  \textsc{Cdae} & 0.391 & 0.523 & 0.418 \\
  \hline\\
\end{tabular}
\end{subtable}

\begin{subtable}[]{\columnwidth}
\vspace{-3mm}
\caption{Netflix}
\vspace{-2mm}
	\centering
\begin{tabular}{ l c c c c }
\hline
 & Recall@20 & Recall@50 & NDCG@100  \\
  \hline

  \textsc{Sw-dae} &\bf 0.370 &\bf 0.458  &\bf 0.404\\
 \textsc{Sw-dae-p}&0.364& 0.453& 0.398\\
  \hdashline
   \textsc{Fc-Prune-50}&0.355&0.445&0.390\\
  \textsc{Fc-reg}&0.355&0.444&0.389\\
  \hline
  \textsc{Mult-vae} &  0.351 &  0.444 &  0.386\\
  \textsc{Mult-dae} & 0.344 & 0.438 & 0.380 \\
  
  {\textsc{Wmf}} & 0.316 & 0.404 & 0.351\\
  \textsc{Slim} & 0.347 & 0.428 & 0.379 \\
  {\textsc{Cdae}} & 0.343 & 0.428 & 0.376 \\
  \hline\\
\end{tabular}
\end{subtable}

\begin{subtable}[]{\columnwidth}
\vspace{-3mm}
\caption{MSD}
\vspace{-2mm}
	\centering
\begin{tabular}{ l c c c c }
\hline
 & Recall@20 & Recall@50 & NDCG@100  \\
  \hline

    \textsc{Sw-dae} &\bf 0.317 &\bf 0.416 & \bf 0.372 \\
  \textsc{Sw-dae-p}&0.313 & 0.413& 0.369\\
  \hdashline
    \textsc{Fc-Prune-50}&0.304&0.397&0.356\\
  \textsc{Fc-reg}&0.300&0.393&0.352\\
  \hline
  \textsc{Mult-vae} &  0.266 &  0.364 &  0.316 \\
  \textsc{Mult-dae}  &  0.266 & 0.363 & 0.313 \\
   
  {\textsc{Wmf}} & 0.211 & 0.312 & 0.257 \\
  \textsc{Slim} & --- & --- & --- \\
  {\textsc{Cdae}} & 0.188 & 0.283 & 0.237\\
  \hline\\
\end{tabular}
\end{subtable}
\label{tab.results}
\vspace{-.5em}
\end{table}

	\begin{figure}
		\includegraphics[scale=0.37]{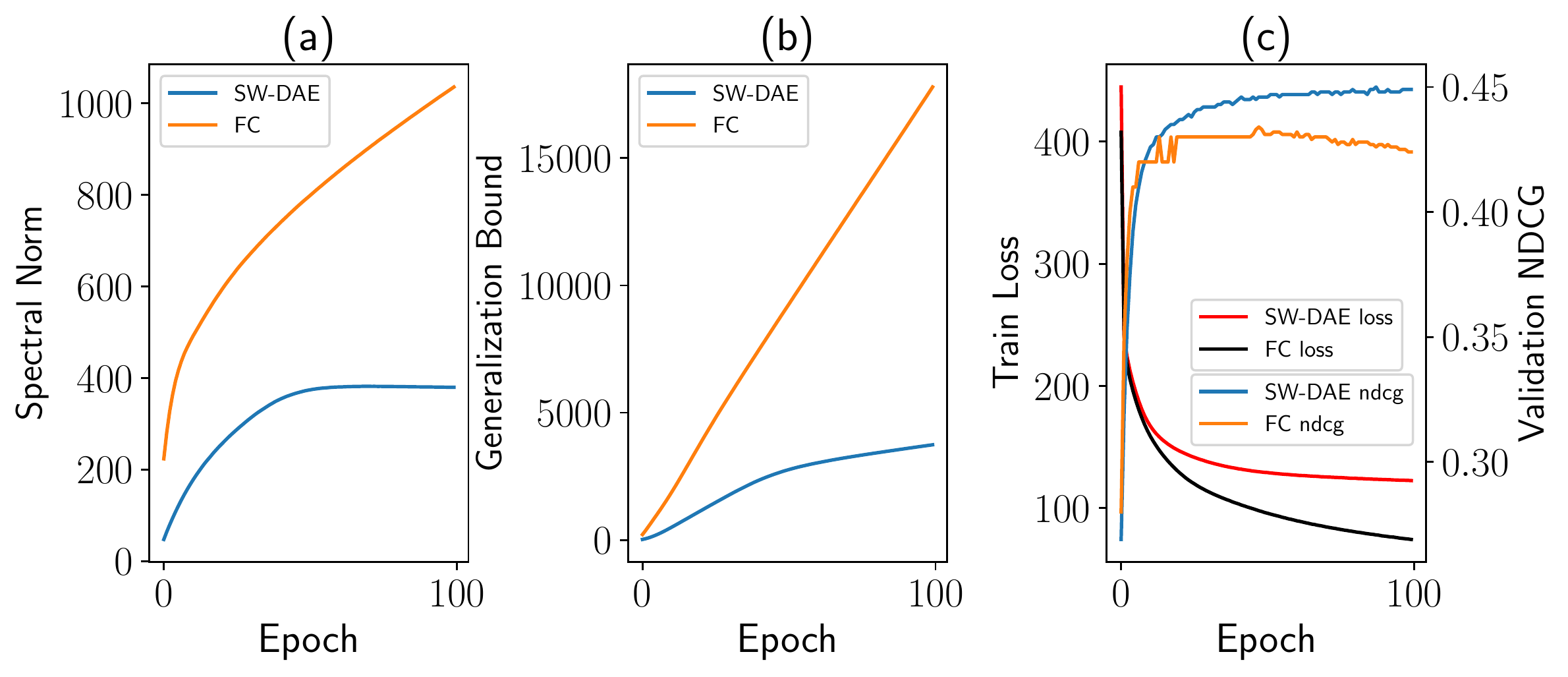}
        \vspace{-7mm}
		\caption{ (a) The spectral norm of \textsc{Sw-dae} and \textsc{Fc-dae}, (b) the generalization error upper bound and (c) the training loss and validation NDCG for the ML20M dataset. \textsc{Sw-dae} has a lower spectral norm and hence a lower generalization error bound. This manifests as a higher validation NDCG even when the training loss is more than \fc.}
		\label{figure.spectralnorm}
		\vspace{-.5em} 
	\end{figure}

\subsection{Is the Gain Due to More Neurons?} 

\begin{figure}
	\centering
		\setlength{\fboxrule}{0pt}
		\fbox{
		\includegraphics[scale=0.25]{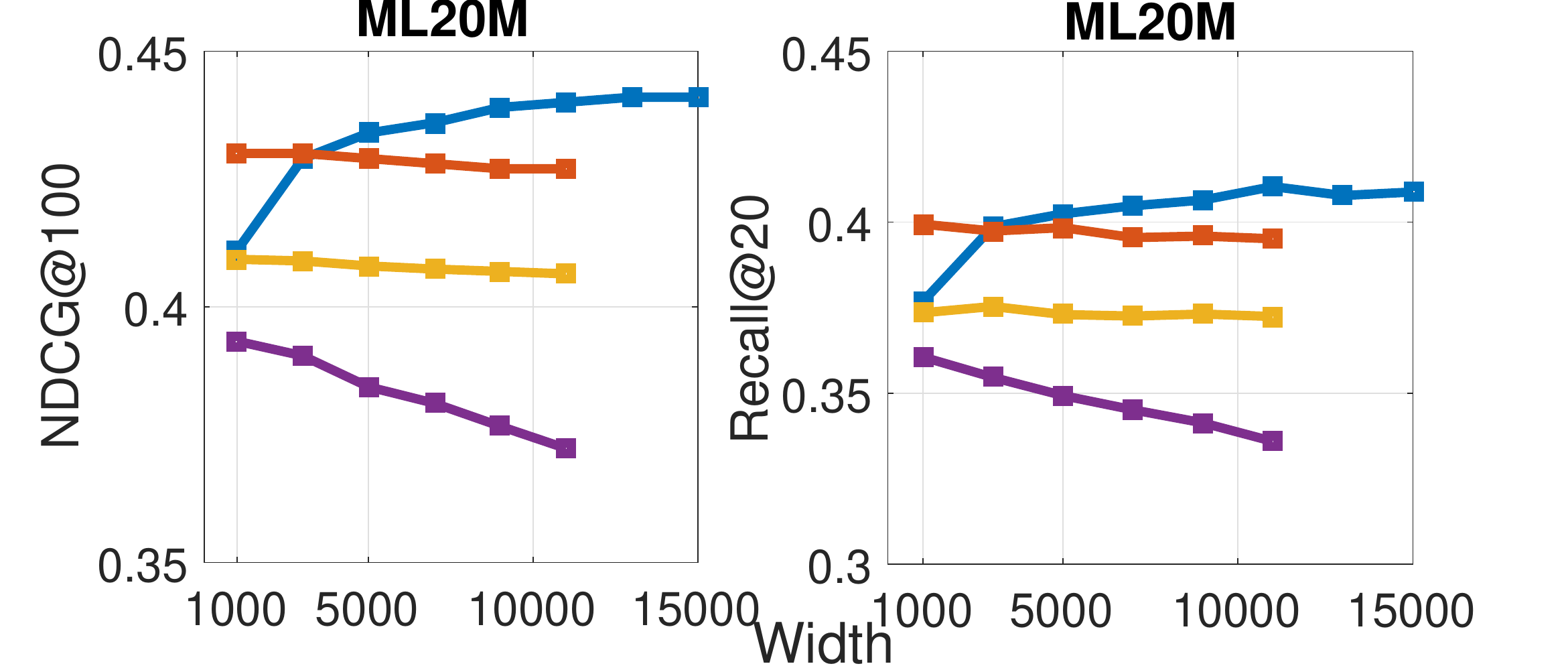}
		}
		\setlength{\fboxrule}{0pt}
		\fbox{
		\includegraphics[scale=0.25]{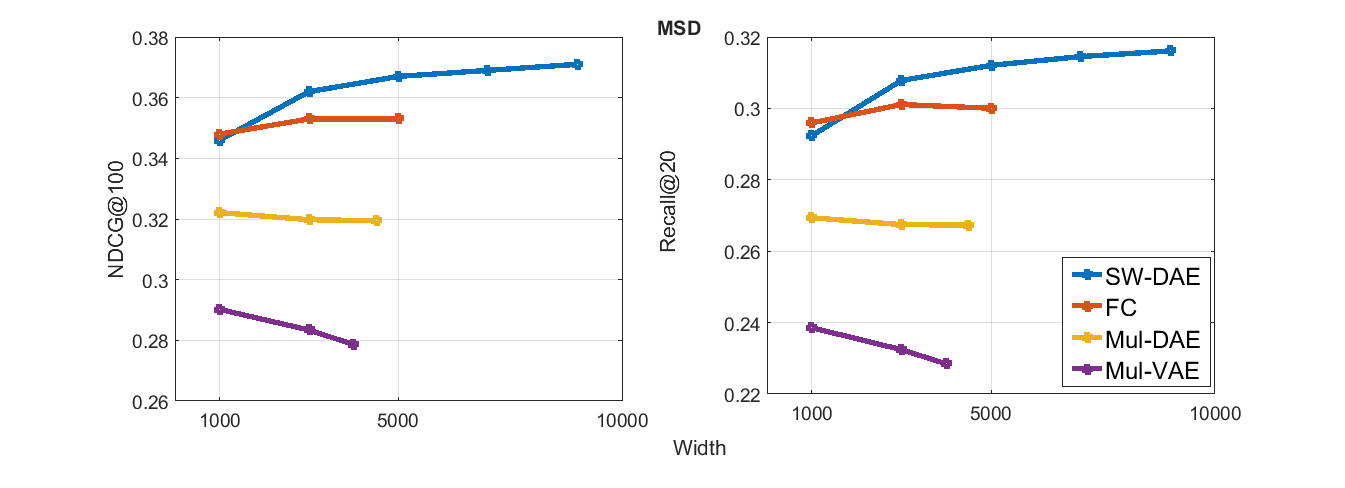}
		}
		\vspace{-7mm}
		\caption{The performance of \textsc{Sw-dae} improves with the width ($K$) of the hidden layer, unlike the baselines, due to better modeling of the item relationships. The performance gain of \textsc{Sw-dae} is due to incorporating the item group information in its structure.}
		\label{figure.width}
		\vspace{-.5em} 
	\end{figure}

We can also make the baseline structures wide and examine their performance in relation to \textsc{Sw-dae}. Figure \ref{figure.width} shows this comparison for the two best baselines \textsc{Mult-dae/vae} on the ML20M and MSD datasets. In addition, we also make a comparison with a \textsc{Fc-dae} that uses the same dropout, activation and loss functions as \textsc{Sw-dae}. All the methods used the same structure of one hidden layer and the width of the models was increased until the GPU memory capacity was reached. 

From Figure \ref{figure.width} we can see that, unlike the baselines, the NDCG@$100$ and Recall@$20$ performance of \textsc{Sw-dae} increases with the number of neurons and finally plateaus. Increasing the number of neurons allows better and more fine-grained modeling of item interactions as each neuron models the interactions of a smaller group of related items. 

We see that for \textsc{Mult-dae/vae} and \textsc{Fc-dae} increasing the number of neurons does not help in performance. In fact the performance of narrower \textsc{Mult-dae/vae} shown in Table \ref{tab.results} was better than their wider counterparts shown in Figure \ref{figure.width}. 

\subsection{Does the Item Group Structure Help?} 
This brings us to the utility of the sparse and wide cluster structure. To isolate the performance gain due to the cluster structure, we can compare the results of \textsc{Sw-dae} and \textsc{Fc-dae} in Figure \ref{figure.width}. Since the corresponding \textsc{Fc-dae} uses the same number of neurons, activation function, and parameter and learning settings as \textsc{Sw-dae}, this improved performance can be attributed to better modeling of item relations as a result of the introduction of the item group structure.

\subsection{Stable Rank and Spectral Norm}
The item group structure removes unnecessary item interaction in the first phase by fixing the structure, and in the second phase \textsc{Sw-dae} only learns useful parameters from the sparse data. Intuitively this can be the rationale for the improved generalization. To understand the generalization ability of \textsc{Sw-dae} more formally, we investigate the spectral norm and stable rank of the weight matrix (either encoder or decoder) of the auto-encoder. 

The stable rank $\srank(\mathbf{W})$ of a matrix $\mathbf{W}$ is defined as $\srank(\mathbf{W})\defeq ||\mathbf{W}||_F^2/||\mathbf{W}||_2^2$. It is the ratio of its squared Frobenius norm and squared spectral norm. Clearly, for any matrix $\mathbf{W}$, we have $1\leq \srank(\mathbf{W})\leq \rank(\mathbf{W})$. The stable rank is more stable than the rank because it is largely unaffected by tiny singular values. More importantly,
it has recently been shown ~\citep[Theorem 1]{neyshabur2017pac} that the generalization error of the neural network depends
on the spectral norm and the stable rank of the weight matrices of the network. Specifically, it can be shown that the generalization error is upper bounded by $O\big(\sqrt{\prod_{j=1}^L||\mathbf{W}_j||_2^2\sum_{j=1}^L \srank(\mathbf{W}_j)/n}\big)$, where $L$ is the number of layers in the network. This upper bound suggests that a smaller spectral norm (or smoother function mapping) and smaller stable rank lead to better generalization. We note that this upper bound depends on the product of the square of the spectral norms of all the layers, therefore, a smaller spectral norm is highly desirable.

We can compare the spectral norm and generalization bound of \sw with the corresponding \fc compare which is identical to \sw in every respect except the fact that \textsc{Fc-dae} does not have the item group structure. In Figure \ref{figure.spectralnorm} (a) we plot the spectral norm of weight matrix for the encoder for \textsc{Sw-dae} and \textsc{Fc-dae}\footnote{Similar results are observed for the decoder and on other datasets.} on the ML20M dataset. We see that \sw has a much lower spectral norm then \fc throughout the training. In Figure \ref{figure.spectralnorm} (b) we plot the generalization error upper bound for both \sw and \fc and again we see that \sw has a much lower generalization error upper bound. We also note that, unlike \fc, by the end of the 100-th epoch the spectral norm and generalization error stabilize for \sw. Therefore, as expected, in Figure \ref{figure.spectralnorm} (c) we see that as the training progresses \sw has a much better validation NDCG even though the training error is higher than \fc. Consequently, the effect of introducing the item group structure is the reduction of the spectral norm of network weight matrices and this results in better generalization.

\subsection{Other Methods of Structure Learning} 
Pruning the network connections after training followed by retraining is one popular way of learning the connectivity structure. In Table \ref{tab.prune} we show the behavior of the pruning method (\textsc{Fc-prune}) on the ML20M dataset.
We used the best FC model based on validation, pruned it following \cite{han2015learning} and then retrained it. As we prune more connections, the network becomes sparser, however, its performance also drops. Prune-90 has the same $10\%$ sparsity level as \textsc{Sw-dae}, but its performance is much lower than \textsc{Sw-dae}.

We can also make the activation of the hidden layer sparse by introducing a $\ell_1$ penalty ($\lambda_1$) on the activation of the hidden layer. Again we use the best \textsc{Fc-dae} based on the validation performance for the experiments and compare its performance with \textsc{Sw-dae}. Table \ref{tab.l1} shows this comparison on the ML20M dataset. \textsc{Sw-dae} is considerably better than any regularized model. We can see that too little or too much regularization are both undesirable and there exists an optimal level in-between (around $\lambda_1$ = $10^{-3}$) that gives the best performance.
\begin{table}[]
\caption{The best FC architecture was chosen for pruning. The reported results are on the ML20M data. Prune-90 has the same sparsity level of $10\%$ as \textsc{Sw-dae} but \textsc{Sw-dae} performs much better.  }
\vspace{-3mm}
\small
\centering
\begin{tabular}{llll}
\hline
{Method}         & {NDCG@100} &{Recall@20} &{Recall@50} \\
\hline
\textsc{Sw-dae}&\bf 0.442&\bf 0.410&\bf 0.549\\
\hdashline
\textsc{Fc-Prune-90}                 & 0.415    & 0.385   & 0.521   \\ 
\textsc{Fc-Prune-80}&0.424&	0.392 &	0.531 \\
\textsc{Fc-Prune-70}&0.427&	0.395 &	0.531\\
\textsc{Fc-Prune-60}&0.431&	0.399 &	0.534\\
\textsc{Fc-Prune-50}&0.430&	0.399 &    0.535\\
\hline
\end{tabular}

\label{tab.prune}
\vspace{-.5em}
\end{table}

\begin{table}[]
\caption{The best FC architecture on ML20M dataset was chosen for regularization. \textsc{Sw-dae} outperforms the sparsely activated regularized versions of DAE.}
\vspace{-3mm}
\small
\centering
\begin{tabular}{llll}
\hline
Method      & NDCG@100 & Recall@20 & Recall@50 \\
\hline
\textsc{Sw-dae}&\bf 0.442&\bf 0.410&\bf 0.549\\
\hdashline
\textsc{Fc-reg} $\lambda_1$ = $10^{-5}$&0.429&	0.396&	0.533\\
\textsc{Fc-reg} $\lambda_1$ = $10^{-4}$&0.430&	0.398&	0.534\\
\textsc{Fc-reg} $\lambda_1$ = $10^{-3}$&0.431&	0.399&	0.535\\
\textsc{Fc-reg} $\lambda_1$ = $10^{-2}$&0.425&	0.395&	0.533\\
\textsc{Fc-reg} $\lambda_1$ = $10^{-1}$&0.376&	0.343&	0.476\\
\hline
\end{tabular}

\label{tab.l1}
\vspace{-.5em}
\end{table}

\begin{table}[]
\caption{The comparison of the running time of the components of two item grouping methods (Algorithm 1 \& 2) in seconds along with their respective accuracies.}
\fontsize{6.5}{7.0}\selectfont
\label{tab:time}
\begin{tabular}{llll|cc}
\hline
\textbf{ML20M}   & {Spectrum} & {K-means} & {Total} & {Recall@50} & {NDCG@100} \\ \hline
Algorithm 1        & 155 sec.                      & 528.6 sec.                  & 683.6 sec.               & 0.549              & 0.442             \\
Algorithm 2    & 41.3 sec.                     & 833.1 sec.                   & 874.4 sec.              & 0.545              & 0.441             \\ \hline
\textbf{Netflix} &                          &                         &                    &                    &                   \\ \hline
Algorithm 1        & 238.5 sec.                    & 1591.1 sec.                  & 1829.6 sec.               & 0.458              & 0.404             \\
Algorithm 2    & 176.8 sec.                    & 400.9 sec.                     &577.7 sec.              & 0.457              & 0.403             \\ \hline
\textbf{MSD}     &                          &                         &                    &                    &                   \\ \hline
Algorithm 1       & 1229.8 sec.                   & 1068.9 sec.                  & 2298.7 sec.              & 0.416              & 0.372             \\
Algorithm 2    & 273.5 sec.                    & 1267.8 sec.                  & 1541.3 sec.             & 0.414              & 0.371             \\ \hline
\end{tabular}
\vspace{-.5em}
\end{table}

\subsection{Analysis of Item Grouping}\label{sec.itemgroupping}
Table \ref{tab:time} shows the comparison of Algorithm \ref{alg.itemcluster} and Algorithm \ref{alg.itemcluster2} in terms of the running time and the recommendation performance. We see that both the algorithms are reasonably fast even on the lager Netflix and MSD datasets. Also, Algorithm \ref{alg.itemcluster2} is faster on all datasets in getting the top $F$ eigenvectors due to operating on the more sparse matrix $\bX$ directly. We also see that the time taken by K-means to cluster the projected items is not a lot. This is due to the items being $F$-dimensional vectors where $F$ is small. As a result, both the algorithms scale well to large and sparse data. Finally, we note that, as expected, Algorithm \ref{alg.itemcluster} provides better accuracy on all three datasets due to discounting for item popularity by normalizing the Laplacian of Equation \ref{equation:laplacian}.  

\subsection{Cold Start}		
We compare the performance of \textsc{Sw-dae} in the cold-start scenario with \textsc{Mult-dae/vae} as they represent the best two baselines. As before we first sample the test users and then we randomly sample 80\% of the events (fold-in set) of the test users for learning their representations and use the remaining 20\% events of the test users for testing. From the test users, we select the cold-start users based on their activity in the fold-in set. The activity is defined by the number of events of each user in the fold-in set. Testing is then done only on these cold-start users. Table \ref{tab.cold} shows this comparison for the ML20M dataset. Consistent with \cite{liang2018variational} we find that \textsc{Multi-vae} performs better than \textsc{Multi-dae} for the user cold-start. However, we see that \textsc{Sw-dae} outperforms  \textsc{Multi-vae}. We conjecture that the addition of cluster structure acts as an additional source of information that enables better performance when the user data is scarce. A similar trend is observed on the other datasets.

\begin{table}
\caption{\textsc{Sw-dae} outperforms the state-of-the-art \textsc{Mult-dae/vae} in the cold-start scenario on the ML20M dataset.} 
\vspace{-3mm}
\small
\centering
\begin{tabular}{ l c c c c }
\hline
    & Recall@20 & Recall@50 & NDCG@100  \\
  \hline
    \textsc{Sw-dae}&\bf 0.460&\bf 0.619&\bf 0.343\\
  \textsc{Mult-vae} & 0.456& 0.617 &  0.337 \\
  \textsc{Mult-dae} & 0.448 & 0.614 & 0.328 \\
  \hline\\
\end{tabular}

\label{tab.cold}
		\vspace{-.8em} 
\end{table}

 	\begin{figure}[]
	\centering
		\includegraphics[width=2.5in, height=1.2in]{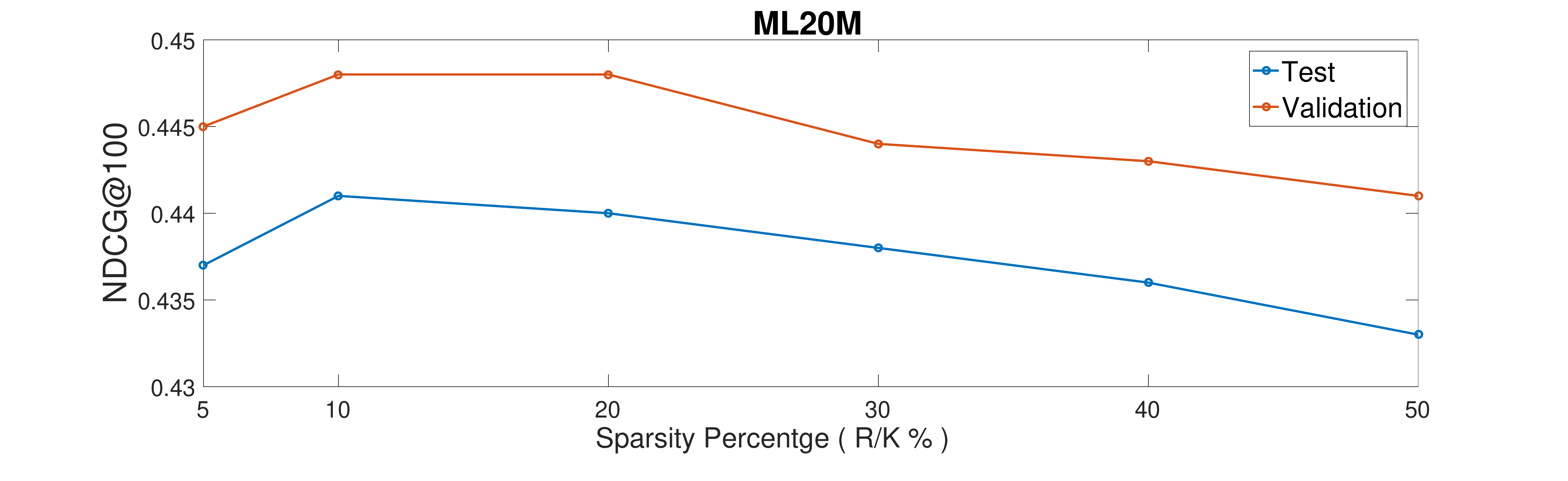}
		\vspace{-3mm}
		\caption{Effect of sparsity on validation and test NDCG@100 on ML20M dataset. X-axis denotes the sparsity level: higher means more dense.}
		\label{figure.effectRndcg}
		\vspace{-.5em} 
	\end{figure}

\subsection{Effect of Sparsity: R/K}
In our experiments, we set the sparsity level to 10\% i.e., $R/K=0.1$ for all the datasets. In this section we empirically investigate the effect of the sparsity level ($R/K$) on the performance of \textsc{Sw-dae.} Figure \ref{figure.effectRndcg} shows the effect of the sparsity level on the validation and test NDCG@100 on ML20M and similar trends were observed for other datasets. The x-axis denotes the sparsity level where larger values denote a more dense network. We can see that the best validation performance was obtained for 10\% and 20\% sparsity. Based on this the 10\% sparsity was chosen for the experiments in the paper. We also note that very sparse and very dense networks both lead to unsatisfactory performance and there exist optimal values of sparsity between 10\% and 20\%.

	\begin{figure}

	\centering
		\setlength{\fboxrule}{0pt}
		\fbox{
		\includegraphics[scale=0.25]{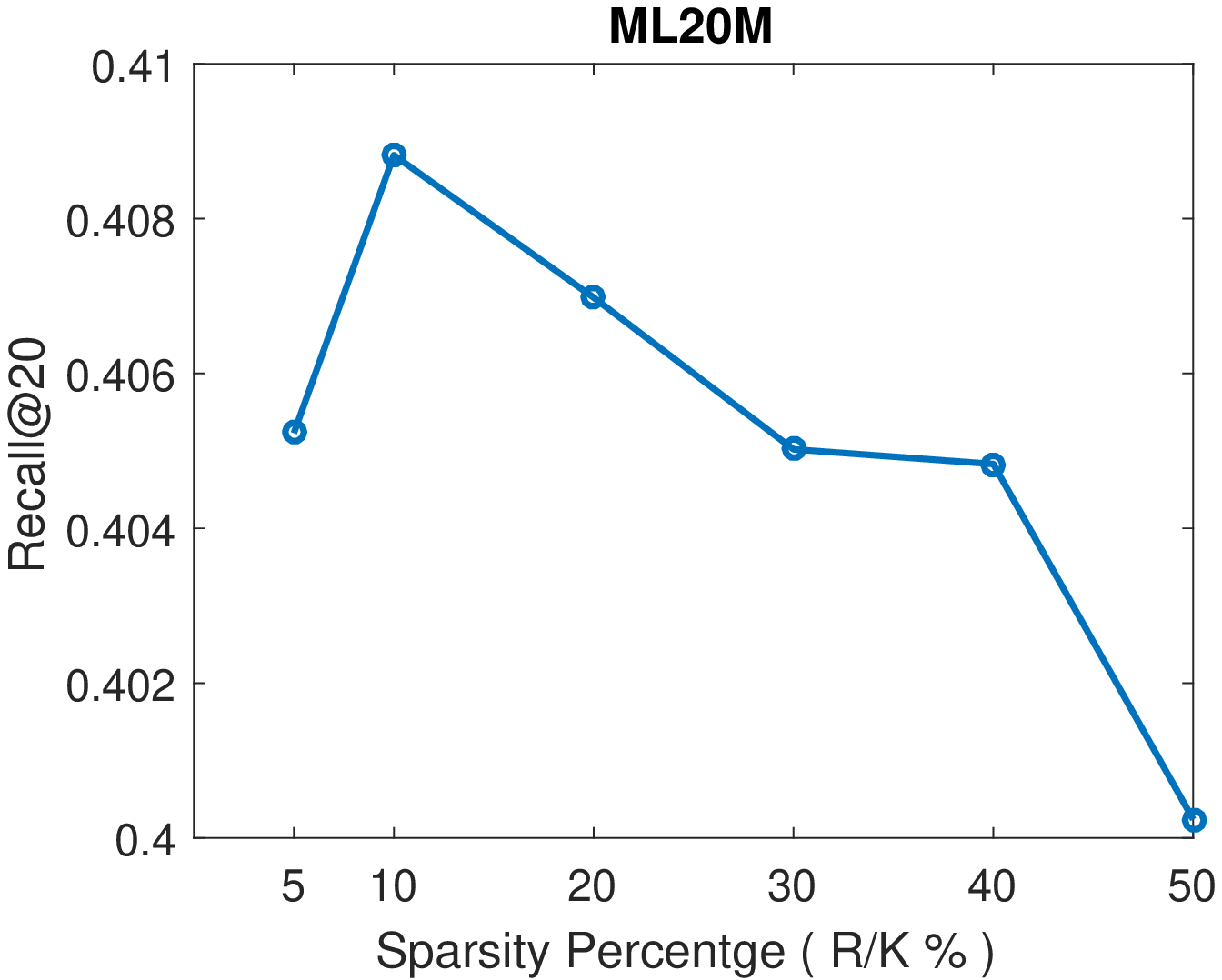}
		}
		\setlength{\fboxrule}{0pt}
		\fbox{
		\includegraphics[scale=0.25]{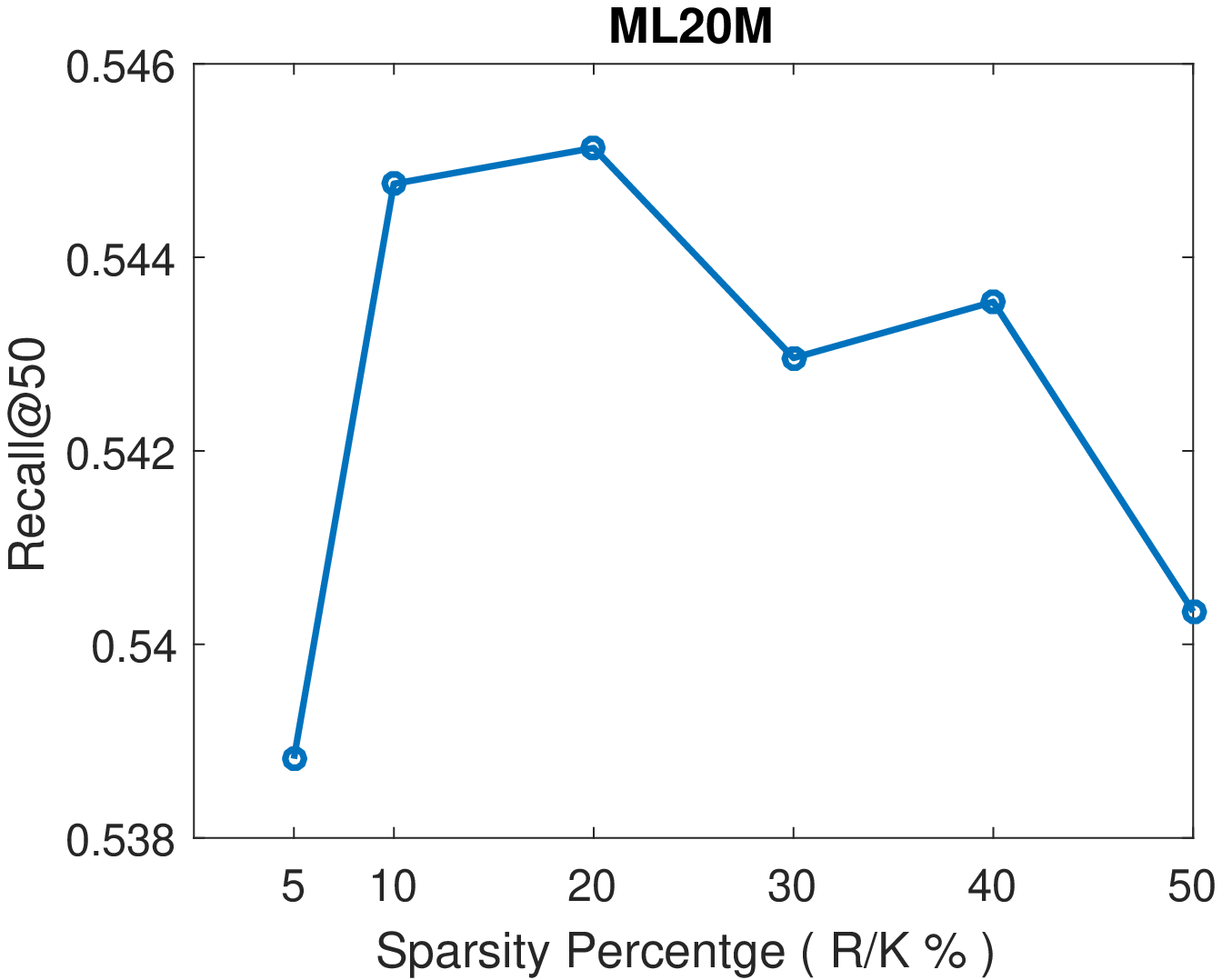}
		}
		\vspace{-3mm}
		\caption{Effect of sparsity on test Recall@20 and Recall@50 on ML20M dataset.}
		\label{figure.effectR}
		\vspace{-.5em}
	\end{figure}	
		
Figure \ref{figure.effectR} shows the effect of sparsity on the test Recall@20 and test Recall@50. We observe a similar trend as the NDCG@100 graph. There exists an optimal level of sparsity between 10\% and 20\% that gives the best performance.

\begin{table}[]
\caption{\textsc{Sw-dae} detects a variety of related items. Below we see four sample neurons that detect documentary, horror, drama, and children genres on the ML20M dataset.}
\vspace{-3mm}
\small
\centering
\begin{tabular}{l|l}
\hline
Neuron ID & Top three genres                               \\
\hline
2785   & Documentary(35\%), Drama(39\%), Comedy(32\%) \\
691    & Horror(73\%), Thriller(37\%), Sci-fi(17\%)   \\
1247   & Drama(88\%), Crime (25\%), Action (25\%)     \\
387    & Children(42\%), Comedy(44\%), Drama(30\%)   \\
\hline
\end{tabular}
\label{tab.qual}
\vspace{-.5em}
\end{table}
\subsection{Qualitative Analysis}
We also examine the resulting \textsc{Sw-dae} structure qualitatively. For this, we obtained the genre information of each movie in the ML20M dataset from IMDB. In Table \ref{tab.qual} we show four sample neurons where each neuron represents an overlapping cluster. The top three genres for each overlapping cluster are shown along with the percentage of movies in the cluster that possess this genre. Each movie can possess more than one genre. We see that neuron 691 primarily detects the genres associated with horror i.e., horror-thriller-scifi movies, similarly, other neurons detect the documentary (neuron 2785), drama(neuron 1247) and children (neuron 387) genres. Thus, the overlapping clusters can pick up related items.

\begin{figure}[]

	\centering
		\setlength{\fboxrule}{0pt}
		\fbox{
		\includegraphics[height=1.1in, width=\columnwidth]{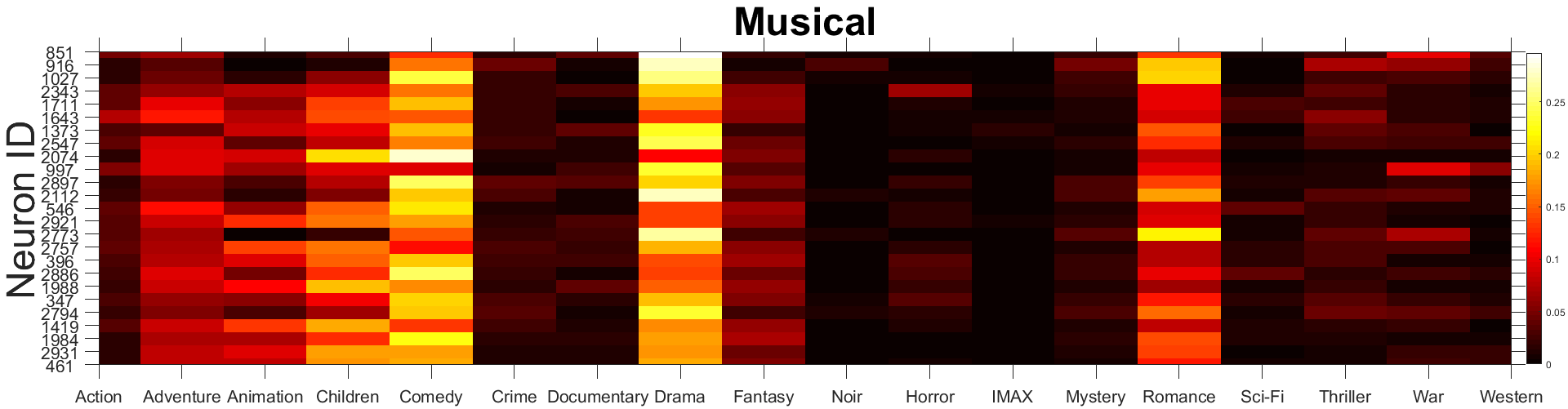}
		}
		\setlength{\fboxrule}{0pt}
		\fbox{
		\includegraphics[height=1.1in, width=\columnwidth]{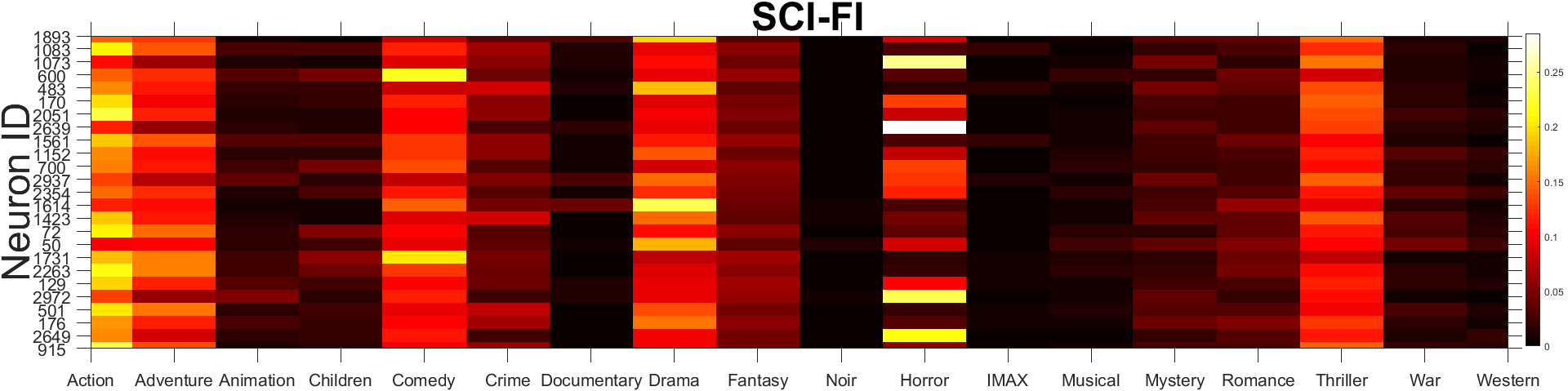}
		}
		\setlength{\fboxrule}{0pt}
		\fbox{
		\includegraphics[height=1.1in, width=\columnwidth]{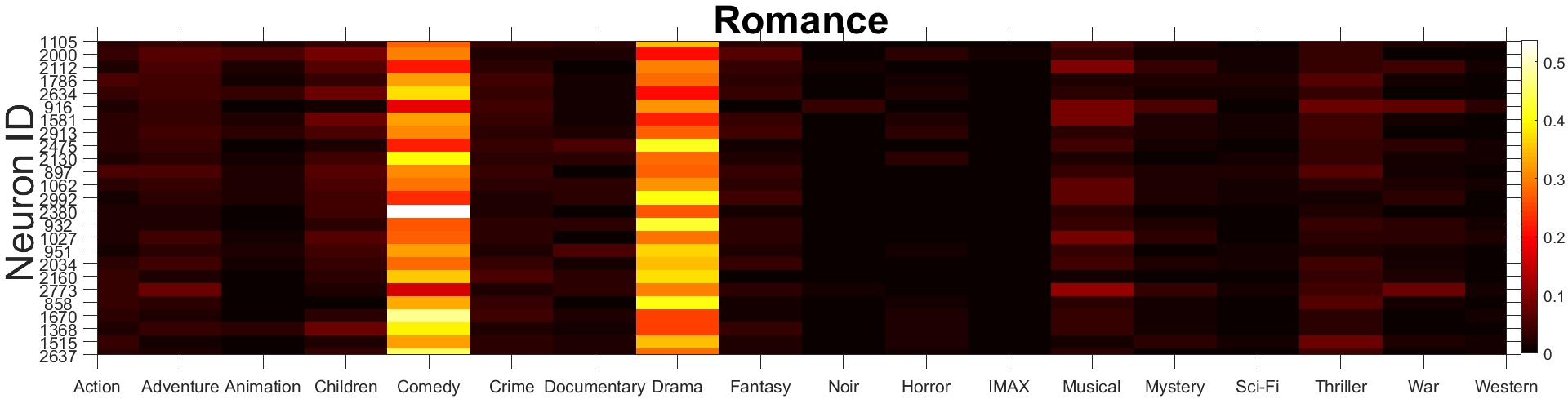}
		}
		\vspace{-6mm}
		\caption{We pick the top 25 neurons possessing a specific genre and then examine the other genres present in the overlapping cluster represented by this genre. The y-axis denotes the neuron ID and the x-axis denotes the other genres. The color denotes the degree to which a genre belongs to a cluster. We can see that the neurons possessing the musical genre also possess a high percentage of comedy, drama, children genres. Similarly, the neurons that possess the romance genres also possess movies with drama and comedy genres but rarely possess the horror, mystery and war genres.}
		\label{figure.heatmap}
		\vspace{-.5em}
	\end{figure}

In Figure \ref{figure.heatmap} we pick the top 25 neurons possessing a specific genre and then examine the other genres present in the overlapping cluster represented by this genre. Specifically, we first select the top 25 neurons with respect to the percentage of movies of a specific genre present in them. Then for each of these neurons, we calculate the percentage of other genres present. Figure \ref{figure.heatmap} shows the heat maps for three genres: musical, romance, and sci-fi. The y-axis represents the neuron IDs and the x-axis represents the various genres. The color denotes the degree to which a genre is present in the corresponding cluster. We can see that the neurons possessing the musical genre also possess a high percentage of comedy, drama, children genres but do not possess the movies with mystery, western, noir genres. Similarly, the neurons that possess the romance genres also possess movies with drama and comedy genres but rarely possess the horror, mystery and war genres. Finally, the sci-fi movies are generally not grouped with IMAX, musical and children movies but are frequently grouped with thriller, horror, action, etc. genres. This shows that the techniques employed for overlapping clustering result in thematically similar overlapping clusters, and movies with complementary genres are grouped together.

\section{Conclusion}
Existing autoencoders use fully-connected layers for making recommendations. They fail to model the tendency of items to be related to only a subset of all items. Rather they connect each neuron to all items and rely on the network to automatically determine which interactions to model. This becomes especially difficult when the data is sparse. To overcome this we proposed the use to structure learning to decide the connectivity pattern of the neurons. We showed that existing structure learning methods can be adopted for recommendation and they outperform the state-of-the-art methods. We then presented a two-stage method that first fixes the structure based on item groups and then trains only the required connections. We conducted extensive experiments to show that our proposed structure learning technique considerably outperforms the baselines, and also performs better in the cold-start scenario. We showed that this improvement is due to having a smaller spectral norm and a lower generalization error upper-bound. Moreover, considerable improvements can be seen even when the same number of flops/parameters are used as the baselines. We also showed that the item grouping phase is fast and scalable for sparse data and the learned overlapping clusters have thematically similar items.  

\section*{Acknowledgment}
Research on this article was supported by Hong Kong Research Grants Council under grants 16202118 and 18200519.
\bibliographystyle{ACM-Reference-Format}
\bibliography{ae}


\begin{thebibliography}{30}


\ifx \showCODEN    \undefined \def \showCODEN     #1{\unskip}     \fi
\ifx \showDOI      \undefined \def \showDOI       #1{#1}\fi
\ifx \showISBNx    \undefined \def \showISBNx     #1{\unskip}     \fi
\ifx \showISBNxiii \undefined \def \showISBNxiii  #1{\unskip}     \fi
\ifx \showISSN     \undefined \def \showISSN      #1{\unskip}     \fi
\ifx \showLCCN     \undefined \def \showLCCN      #1{\unskip}     \fi
\ifx \shownote     \undefined \def \shownote      #1{#1}          \fi
\ifx \showarticletitle \undefined \def \showarticletitle #1{#1}   \fi
\ifx \showURL      \undefined \def \showURL       {\relax}        \fi
\providecommand\bibfield[2]{#2}
\providecommand\bibinfo[2]{#2}
\providecommand\natexlab[1]{#1}
\providecommand\showeprint[2][]{arXiv:#2}

\bibitem[\protect\citeauthoryear{Aggarwal}{Aggarwal}{2016}]%
        {Aggarwal:2016:RST:2931100}
\bibfield{author}{\bibinfo{person}{Charu~C. Aggarwal}.}
  \bibinfo{year}{2016}\natexlab{}.
\newblock \bibinfo{booktitle}{\emph{Recommender Systems: The Textbook}
  (\bibinfo{edition}{1st} ed.)}.
\newblock \bibinfo{publisher}{Springer Publishing Company, Incorporated}.
\newblock
\showISBNx{3319296574, 9783319296579}


\bibitem[\protect\citeauthoryear{Amatriain, Jaimes, Oliver, and
  Pujol}{Amatriain et~al\mbox{.}}{2015}]%
        {Amatriain2015}
\bibfield{author}{\bibinfo{person}{Xavier Amatriain},
  \bibinfo{person}{Alejandro Jaimes}, \bibinfo{person}{Nuria Oliver}, {and}
  \bibinfo{person}{Josep~M. Pujol}.} \bibinfo{year}{2015}\natexlab{}.
\newblock \bibinfo{booktitle}{\emph{Recommender Systems Handbook}}.
\newblock \bibinfo{publisher}{Springer US}, \bibinfo{address}{Boston, MA},
  Chapter Data Mining Methods for Recommender Systems.
\newblock


\bibitem[\protect\citeauthoryear{Baglama and Reichel}{Baglama and
  Reichel}{2005}]%
        {baglama2005augmented}
\bibfield{author}{\bibinfo{person}{James Baglama} {and} \bibinfo{person}{Lothar
  Reichel}.} \bibinfo{year}{2005}\natexlab{}.
\newblock \showarticletitle{Augmented implicitly restarted Lanczos
  bidiagonalization methods}.
\newblock \bibinfo{journal}{\emph{SIAM Journal on Scientific Computing}}
  \bibinfo{volume}{27}, \bibinfo{number}{1} (\bibinfo{year}{2005}),
  \bibinfo{pages}{19--42}.
\newblock


\bibitem[\protect\citeauthoryear{Goodfellow, Bengio, and Courville}{Goodfellow
  et~al\mbox{.}}{2016}]%
        {Goodfellow-et-al-2016}
\bibfield{author}{\bibinfo{person}{Ian Goodfellow}, \bibinfo{person}{Yoshua
  Bengio}, {and} \bibinfo{person}{Aaron Courville}.}
  \bibinfo{year}{2016}\natexlab{}.
\newblock \bibinfo{booktitle}{\emph{Deep Learning}}.
\newblock \bibinfo{publisher}{MIT Press}.
\newblock
\newblock
\shownote{\url{http://www.deeplearningbook.org}.}


\bibitem[\protect\citeauthoryear{Han, Pool, Tran, and Dally}{Han
  et~al\mbox{.}}{2015}]%
        {han2015learning}
\bibfield{author}{\bibinfo{person}{Song Han}, \bibinfo{person}{Jeff Pool},
  \bibinfo{person}{John Tran}, {and} \bibinfo{person}{William Dally}.}
  \bibinfo{year}{2015}\natexlab{}.
\newblock \showarticletitle{Learning both weights and connections for efficient
  neural network}. In \bibinfo{booktitle}{\emph{Advances in neural information
  processing systems}}. \bibinfo{pages}{1135--1143}.
\newblock


\bibitem[\protect\citeauthoryear{He, Liao, Zhang, Nie, Hu, and Chua}{He
  et~al\mbox{.}}{2017}]%
        {he2017neural}
\bibfield{author}{\bibinfo{person}{Xiangnan He}, \bibinfo{person}{Lizi Liao},
  \bibinfo{person}{Hanwang Zhang}, \bibinfo{person}{Liqiang Nie},
  \bibinfo{person}{Xia Hu}, {and} \bibinfo{person}{Tat-Seng Chua}.}
  \bibinfo{year}{2017}\natexlab{}.
\newblock \showarticletitle{Neural collaborative filtering}. In
  \bibinfo{booktitle}{\emph{Proceedings of the 26th International Conference on
  World Wide Web}}. International World Wide Web Conferences Steering
  Committee, \bibinfo{pages}{173--182}.
\newblock


\bibitem[\protect\citeauthoryear{Heckel, Vlachos, Parnell, and
  D{\"u}nner}{Heckel et~al\mbox{.}}{2017}]%
        {heckel2017scalable}
\bibfield{author}{\bibinfo{person}{Reinhard Heckel}, \bibinfo{person}{Michail
  Vlachos}, \bibinfo{person}{Thomas Parnell}, {and} \bibinfo{person}{Celestine
  D{\"u}nner}.} \bibinfo{year}{2017}\natexlab{}.
\newblock \showarticletitle{Scalable and interpretable product recommendations
  via overlapping co-clustering}. In \bibinfo{booktitle}{\emph{Data Engineering
  (ICDE), 2017 IEEE 33rd International Conference on}}. IEEE,
  \bibinfo{pages}{1033--1044}.
\newblock


\bibitem[\protect\citeauthoryear{Hu, Koren, and Volinsky}{Hu
  et~al\mbox{.}}{2008}]%
        {Hu:2008:CFI:1510528.1511352}
\bibfield{author}{\bibinfo{person}{Yifan Hu}, \bibinfo{person}{Yehuda Koren},
  {and} \bibinfo{person}{Chris Volinsky}.} \bibinfo{year}{2008}\natexlab{}.
\newblock \showarticletitle{Collaborative Filtering for Implicit Feedback
  Datasets}. In \bibinfo{booktitle}{\emph{Proceedings of the 2008 Eighth IEEE
  International Conference on Data Mining}} \emph{(\bibinfo{series}{ICDM
  '08})}. \bibinfo{publisher}{IEEE Computer Society},
  \bibinfo{address}{Washington, DC, USA}, \bibinfo{pages}{263--272}.
\newblock
\showISBNx{978-0-7695-3502-9}
\urldef\tempurl%
\url{https://doi.org/10.1109/ICDM.2008.22}
\showDOI{\tempurl}


\bibitem[\protect\citeauthoryear{Jiang, Zhang, Zhang, and Xiao}{Jiang
  et~al\mbox{.}}{2013}]%
        {Jiang2013ANS}
\bibfield{author}{\bibinfo{person}{Xiaojuan Jiang}, \bibinfo{person}{Yinghua
  Zhang}, \bibinfo{person}{Wensheng Zhang}, {and} \bibinfo{person}{Xian Xiao}.}
  \bibinfo{year}{2013}\natexlab{}.
\newblock \showarticletitle{A novel sparse auto-encoder for deep unsupervised
  learning}.
\newblock \bibinfo{journal}{\emph{2013 Sixth International Conference on
  Advanced Computational Intelligence (ICACI)}} (\bibinfo{year}{2013}),
  \bibinfo{pages}{256--261}.
\newblock


\bibitem[\protect\citeauthoryear{Khawar and Zhang}{Khawar and Zhang}{2019a}]%
        {khawar2019conformative}
\bibfield{author}{\bibinfo{person}{Farhan Khawar} {and}
  \bibinfo{person}{Nevin~L Zhang}.} \bibinfo{year}{2019}\natexlab{a}.
\newblock \showarticletitle{Conformative filtering for implicit feedback data}.
  In \bibinfo{booktitle}{\emph{European Conference on Information Retrieval}}.
  Springer, \bibinfo{pages}{164--178}.
\newblock


\bibitem[\protect\citeauthoryear{Khawar and Zhang}{Khawar and Zhang}{2019b}]%
        {khawar2019modeling}
\bibfield{author}{\bibinfo{person}{Farhan Khawar} {and}
  \bibinfo{person}{Nevin~L Zhang}.} \bibinfo{year}{2019}\natexlab{b}.
\newblock \showarticletitle{Modeling Multidimensional User Preferences for
  Collaborative Filtering}. In \bibinfo{booktitle}{\emph{2019 IEEE 35th
  International Conference on Data Engineering (ICDE)}}. IEEE,
  \bibinfo{pages}{1618--1621}.
\newblock


\bibitem[\protect\citeauthoryear{Khoshneshin and Street}{Khoshneshin and
  Street}{2010}]%
        {khoshneshin2010incremental}
\bibfield{author}{\bibinfo{person}{Mohammad Khoshneshin} {and}
  \bibinfo{person}{W~Nick Street}.} \bibinfo{year}{2010}\natexlab{}.
\newblock \showarticletitle{Incremental collaborative filtering via
  evolutionary co-clustering}. In \bibinfo{booktitle}{\emph{Proceedings of the
  fourth ACM conference on Recommender systems}}. ACM,
  \bibinfo{pages}{325--328}.
\newblock


\bibitem[\protect\citeauthoryear{Kingma and Ba}{Kingma and Ba}{2014}]%
        {kingma2014adam}
\bibfield{author}{\bibinfo{person}{Diederik~P Kingma} {and}
  \bibinfo{person}{Jimmy Ba}.} \bibinfo{year}{2014}\natexlab{}.
\newblock \showarticletitle{Adam: A method for stochastic optimization}.
\newblock \bibinfo{journal}{\emph{arXiv preprint arXiv:1412.6980}}
  (\bibinfo{year}{2014}).
\newblock


\bibitem[\protect\citeauthoryear{Koren and Bell}{Koren and Bell}{2015}]%
        {Koren2015}
\bibfield{author}{\bibinfo{person}{Yehuda Koren} {and} \bibinfo{person}{Robert
  Bell}.} \bibinfo{year}{2015}\natexlab{}.
\newblock \bibinfo{booktitle}{\emph{Recommender Systems Handbook}}.
\newblock \bibinfo{publisher}{Springer US}, \bibinfo{address}{Boston, MA},
  Chapter Advances in Collaborative Filterin, \bibinfo{pages}{77--118}.
\newblock


\bibitem[\protect\citeauthoryear{Lehoucq and Sorensen}{Lehoucq and
  Sorensen}{2000}]%
        {cbai2000templates}
\bibfield{author}{\bibinfo{person}{R. Lehoucq} {and} \bibinfo{person}{D.
  Sorensen}.} \bibinfo{year}{2000}\natexlab{}.
\newblock \bibinfo{booktitle}{\emph{Templates for the solution of algebraic
  eigenvalue problems: a practical guide}}.
\newblock \bibinfo{publisher}{SIAM}, \bibinfo{address}{Philadelphia}, Chapter
  Implicitly Restarted Lanczos Method (Section 4.5).
\newblock


\bibitem[\protect\citeauthoryear{Liang, Krishnan, Hoffman, and Jebara}{Liang
  et~al\mbox{.}}{2018}]%
        {liang2018variational}
\bibfield{author}{\bibinfo{person}{Dawen Liang}, \bibinfo{person}{Rahul~G
  Krishnan}, \bibinfo{person}{Matthew~D Hoffman}, {and} \bibinfo{person}{Tony
  Jebara}.} \bibinfo{year}{2018}\natexlab{}.
\newblock \showarticletitle{Variational Autoencoders for Collaborative
  Filtering}. In \bibinfo{booktitle}{\emph{Proceedings of the 2018 World Wide
  Web Conference on World Wide Web}}. International World Wide Web Conferences
  Steering Committee, \bibinfo{pages}{689--698}.
\newblock


\bibitem[\protect\citeauthoryear{Marlin}{Marlin}{2004}]%
        {marlin2004collaborative}
\bibfield{author}{\bibinfo{person}{Benjamin Marlin}.}
  \bibinfo{year}{2004}\natexlab{}.
\newblock \bibinfo{booktitle}{\emph{Collaborative filtering: A machine learning
  perspective}}.
\newblock \bibinfo{publisher}{University of Toronto}.
\newblock


\bibitem[\protect\citeauthoryear{McFee, Bertin-Mahieux, Ellis, and
  Lanckriet}{McFee et~al\mbox{.}}{2012}]%
        {mcfee2012million}
\bibfield{author}{\bibinfo{person}{Brian McFee}, \bibinfo{person}{Thierry
  Bertin-Mahieux}, \bibinfo{person}{Daniel~PW Ellis}, {and}
  \bibinfo{person}{Gert~RG Lanckriet}.} \bibinfo{year}{2012}\natexlab{}.
\newblock \showarticletitle{The million song dataset challenge}. In
  \bibinfo{booktitle}{\emph{Proceedings of the 21st International Conference on
  World Wide Web}}. ACM, \bibinfo{pages}{909--916}.
\newblock


\bibitem[\protect\citeauthoryear{Neyshabur, Bhojanapalli, and Srebro}{Neyshabur
  et~al\mbox{.}}{2017}]%
        {neyshabur2017pac}
\bibfield{author}{\bibinfo{person}{Behnam Neyshabur}, \bibinfo{person}{Srinadh
  Bhojanapalli}, {and} \bibinfo{person}{Nathan Srebro}.}
  \bibinfo{year}{2017}\natexlab{}.
\newblock \showarticletitle{A pac-bayesian approach to spectrally-normalized
  margin bounds for neural networks}.
\newblock \bibinfo{journal}{\emph{arXiv preprint arXiv:1707.09564}}
  (\bibinfo{year}{2017}).
\newblock


\bibitem[\protect\citeauthoryear{Ng, Jordan, and Weiss}{Ng
  et~al\mbox{.}}{2002}]%
        {ng2002spectral}
\bibfield{author}{\bibinfo{person}{Andrew~Y Ng}, \bibinfo{person}{Michael~I
  Jordan}, {and} \bibinfo{person}{Yair Weiss}.}
  \bibinfo{year}{2002}\natexlab{}.
\newblock \showarticletitle{On spectral clustering: Analysis and an algorithm}.
  In \bibinfo{booktitle}{\emph{Advances in neural information processing
  systems}}. \bibinfo{pages}{849--856}.
\newblock


\bibitem[\protect\citeauthoryear{Ning and Karypis}{Ning and Karypis}{2011}]%
        {ning2011slim}
\bibfield{author}{\bibinfo{person}{Xia Ning} {and} \bibinfo{person}{George
  Karypis}.} \bibinfo{year}{2011}\natexlab{}.
\newblock \showarticletitle{Slim: Sparse linear methods for top-n recommender
  systems}. In \bibinfo{booktitle}{\emph{Data Mining (ICDM), 2011 IEEE 11th
  International Conference on}}. IEEE, \bibinfo{pages}{497--506}.
\newblock


\bibitem[\protect\citeauthoryear{Osting, White, and Oudet}{Osting
  et~al\mbox{.}}{2014}]%
        {osting2014minimal}
\bibfield{author}{\bibinfo{person}{Braxton Osting}, \bibinfo{person}{Chris~D
  White}, {and} \bibinfo{person}{{\'E}douard Oudet}.}
  \bibinfo{year}{2014}\natexlab{}.
\newblock \showarticletitle{Minimal Dirichlet energy partitions for graphs}.
\newblock \bibinfo{journal}{\emph{SIAM Journal on Scientific Computing}}
  \bibinfo{volume}{36}, \bibinfo{number}{4} (\bibinfo{year}{2014}),
  \bibinfo{pages}{A1635--A1651}.
\newblock


\bibitem[\protect\citeauthoryear{O’Connor and Herlocker}{O’Connor and
  Herlocker}{1999}]%
        {o1999clustering}
\bibfield{author}{\bibinfo{person}{Mark O’Connor} {and} \bibinfo{person}{Jon
  Herlocker}.} \bibinfo{year}{1999}\natexlab{}.
\newblock \showarticletitle{Clustering items for collaborative filtering}. In
  \bibinfo{booktitle}{\emph{Proceedings of the ACM SIGIR workshop on
  recommender systems}}, Vol.~\bibinfo{volume}{128}. UC Berkeley.
\newblock


\bibitem[\protect\citeauthoryear{Sarwar, Karypis, Konstan, and Riedl}{Sarwar
  et~al\mbox{.}}{2002}]%
        {sarwar2002recommender}
\bibfield{author}{\bibinfo{person}{Badrul~M Sarwar}, \bibinfo{person}{George
  Karypis}, \bibinfo{person}{Joseph Konstan}, {and} \bibinfo{person}{John
  Riedl}.} \bibinfo{year}{2002}\natexlab{}.
\newblock \showarticletitle{Recommender systems for large-scale e-commerce:
  Scalable neighborhood formation using clustering}. In
  \bibinfo{booktitle}{\emph{Proceedings of the fifth international conference
  on computer and information technology}}, Vol.~\bibinfo{volume}{1}.
  \bibinfo{pages}{291--324}.
\newblock


\bibitem[\protect\citeauthoryear{Ungar and Foster}{Ungar and Foster}{1998}]%
        {ungar1998clustering}
\bibfield{author}{\bibinfo{person}{Lyle~H Ungar} {and} \bibinfo{person}{Dean~P
  Foster}.} \bibinfo{year}{1998}\natexlab{}.
\newblock \showarticletitle{Clustering methods for collaborative filtering}. In
  \bibinfo{booktitle}{\emph{AAAI workshop on recommendation systems}},
  Vol.~\bibinfo{volume}{1}. \bibinfo{pages}{114--129}.
\newblock


\bibitem[\protect\citeauthoryear{Vincent, Larochelle, Bengio, and
  Manzagol}{Vincent et~al\mbox{.}}{2008}]%
        {vincent2008extracting}
\bibfield{author}{\bibinfo{person}{Pascal Vincent}, \bibinfo{person}{Hugo
  Larochelle}, \bibinfo{person}{Yoshua Bengio}, {and}
  \bibinfo{person}{Pierre-Antoine Manzagol}.} \bibinfo{year}{2008}\natexlab{}.
\newblock \showarticletitle{Extracting and composing robust features with
  denoising autoencoders}. In \bibinfo{booktitle}{\emph{Proceedings of the 25th
  international conference on Machine learning}}. ACM,
  \bibinfo{pages}{1096--1103}.
\newblock


\bibitem[\protect\citeauthoryear{Wu, DuBois, Zheng, and Ester}{Wu
  et~al\mbox{.}}{2016a}]%
        {wu2016collaborative}
\bibfield{author}{\bibinfo{person}{Yao Wu}, \bibinfo{person}{Christopher
  DuBois}, \bibinfo{person}{Alice~X Zheng}, {and} \bibinfo{person}{Martin
  Ester}.} \bibinfo{year}{2016}\natexlab{a}.
\newblock \showarticletitle{Collaborative denoising auto-encoders for top-n
  recommender systems}. In \bibinfo{booktitle}{\emph{Proceedings of the Ninth
  ACM International Conference on Web Search and Data Mining}}. ACM,
  \bibinfo{pages}{153--162}.
\newblock


\bibitem[\protect\citeauthoryear{Wu, Liu, Xie, Ester, and Yang}{Wu
  et~al\mbox{.}}{2016b}]%
        {Wu:2016:CIC:2835776.2835836}
\bibfield{author}{\bibinfo{person}{Yao Wu}, \bibinfo{person}{Xudong Liu},
  \bibinfo{person}{Min Xie}, \bibinfo{person}{Martin Ester}, {and}
  \bibinfo{person}{Qing Yang}.} \bibinfo{year}{2016}\natexlab{b}.
\newblock \showarticletitle{CCCF: Improving Collaborative Filtering via
  Scalable User-Item Co-Clustering}. In \bibinfo{booktitle}{\emph{Proceedings
  of the Ninth ACM International Conference on Web Search and Data Mining}}
  \emph{(\bibinfo{series}{WSDM '16})}. \bibinfo{publisher}{ACM},
  \bibinfo{address}{New York, NY, USA}, \bibinfo{pages}{73--82}.
\newblock
\showISBNx{978-1-4503-3716-8}
\urldef\tempurl%
\url{https://doi.org/10.1145/2835776.2835836}
\showDOI{\tempurl}


\bibitem[\protect\citeauthoryear{Yu, Ren, Sun, Gu, Sturt, Khandelwal, Norick,
  and Han}{Yu et~al\mbox{.}}{2014}]%
        {yu2014personalized}
\bibfield{author}{\bibinfo{person}{Xiao Yu}, \bibinfo{person}{Xiang Ren},
  \bibinfo{person}{Yizhou Sun}, \bibinfo{person}{Quanquan Gu},
  \bibinfo{person}{Bradley Sturt}, \bibinfo{person}{Urvashi Khandelwal},
  \bibinfo{person}{Brandon Norick}, {and} \bibinfo{person}{Jiawei Han}.}
  \bibinfo{year}{2014}\natexlab{}.
\newblock \showarticletitle{Personalized entity recommendation: A heterogeneous
  information network approach}. In \bibinfo{booktitle}{\emph{Proceedings of
  the 7th ACM international conference on Web search and data mining}}. ACM,
  \bibinfo{pages}{283--292}.
\newblock


\bibitem[\protect\citeauthoryear{Zosso, Osting, and Osher}{Zosso
  et~al\mbox{.}}{2015}]%
        {zosso2015dirichlet}
\bibfield{author}{\bibinfo{person}{Dominique Zosso}, \bibinfo{person}{Braxton
  Osting}, {and} \bibinfo{person}{Stanley~J Osher}.}
  \bibinfo{year}{2015}\natexlab{}.
\newblock \showarticletitle{A dirichlet energy criterion for graph-based image
  segmentation}. In \bibinfo{booktitle}{\emph{2015 IEEE International
  Conference on Data Mining Workshop (ICDMW)}}. IEEE,
  \bibinfo{pages}{821--830}.
\newblock


\end{thebibliography}
\end{document}